\documentclass[preprint]{aastex}

\input psfig
\begin{document}

\title{Does the fine-structure constant vary with cosmological epoch?
}

\author{John N. Bahcall}
\affil{Institute for Advanced Study, Olden Lane, Princeton, NJ 08540}
\author{Charles L. Steinhardt and David Schlegel}
\affil{Department of Astrophysical Sciences, Princeton University, Princeton, NJ 08544}

\begin{abstract}
We use the strong nebular emission lines of O~{\sc iii}, 5007~\AA\ and 4959~\AA, to set a robust upper
limit on the time dependence of the fine structure constant. We find $|\alpha^{-1} d \, \alpha(t)/dt |
~<~ 2\times 10^{-13}$ yr$^{-1}$, corresponding to $ \Delta \alpha/{\alpha(0)}~=~\left(0.7 \pm 1.4\right)
\times 10^{-4}$ for quasars with $0.16 < z < 0.80$  obtained from the SDSS Early Data Release.  Using a
blind analysis, we show that the upper limit given here is invariant with respect to 17 different ways of
selecting the sample and analyzing the data. As a by-product, we show that the ratio of transition
probabilities corresponding to  the 5007~\AA\ and \hbox{4959~\AA\ } lines is $2.99 \pm 0.02$, in good
agreement with (but more accurate than) theoretical estimates. We compare and
 contrast the O~{\sc iii} emission
line method used here with the Many-Multiplet method that was used recently to suggest evidence for a
time-dependent $\alpha$. In an Appendix, we analyze the quasars from the recently available SDSS Data
Release One sample and find $\Delta \alpha/{\alpha(0)}~=~\left(1.2 \pm 0.7\right) \times 10^{-4}$.

\end{abstract}


\section{Introduction}
\label{sec:introduction}

We develop in this paper a robust analysis strategy for using the
fine-structure
splitting of the O~{\sc iii}
emission lines of distant quasars or galaxies to test whether the fine-structure constant,
$\alpha$, depends upon cosmic time.  We concentrate here on the two strongest nebular emission lines of
O~{\sc iii}, 5007~\AA\ and 4959~\AA, which were first used for this purpose by Bahcall \& Salpeter
(1965).\nocite{bahcall65}

We begin this introduction with a brief history of astronomical studies of the time dependence of the
fine structure constant (\S~\ref{subsec:history}) and then describe some of the relative advantages
of using either absorption lines or emission lines for this purpose
(\S~\ref{subsec:absorptionversusemission}). Next, we provide an outline of the paper
(\S~\ref{subsec:outlineofpaper}). Finally, since some of the material in this paper is intended for
experts, we provide suggestions as to how the paper should be read, or not read
(\S~\ref{subsec:howread}).

We limit the discussion in this paper to tests in which $\alpha$ is the only dimensionless fundamental
constant whose time (or space) dependence can affect the measurements of quasar spectra in any
significant way. For simplicity, we describe our results as if $\alpha$ could only change in time, not in
space. However, our experimental results may be interpreted as placing constraints on the variation of
coupling constants in the context of a more general theoretical framework.

\subsection{A brief history}
\label{subsec:history}

Savedoff (1956)\nocite{savedoff} set the first astronomical upper limit on the time dependence of
$\alpha$ using the fine-structure splitting of emission lines of
N~{\sc ii} and Ne~{\sc iii} in the spectrum of the
nearby radio galaxy Cygnus A. He reported $\alpha^2(z = 0.057)/\alpha^2(0)~=~1.0036 \pm 0.003$. This
technique was first applied to distant quasar spectra by Bahcall \& Salpeter (1965), who used emission
lines of O~{\sc iii} and Ne~{\sc iii} in the spectra of 3C~47 and 3C~147 to set an upper limit of $|\alpha^{-1} d \,
\alpha(t)/dt | ~<~ 10^{-12}$ yr$^{-1}$. Bahcall \& Schmidt (1967)\nocite{bahcall67b} obtained a similar
limit, $<~ 10^{-12}$ yr$^{-1}$,  using the O~{\sc iii} emission lines in five radio galaxies. Here, and
elsewhere in the paper, we shall compare limits on the time-dependence of $\alpha$ obtained at different
cosmic epochs by assuming $\alpha$ depends linearly on cosmic time for redshifts less than five. The
assumption of linearity may misrepresent the actual time dependence, if any,  but it provides a simple
way of comparing measurements at different redshifts.

The technique of using emission lines to investigate the time dependence of $\alpha$ was abandoned after
1967. In this sense, the present paper is an attempt to push the frontiers of the subject backwards three
and a half decades.

Bahcall, Sargent, \& Schmidt~(1967)\nocite{bahcall67a} first used quasar absorption (not emission) lines
to investigate the time dependence of $\alpha$.  They used the doublet
fine-structure splitting of Si~{\sc ii}
and Si~{\sc iv} absorption lines in the spectrum of the bright radio quasar 3C~191, $z~=~1.95$. Although the
results referred to a relatively large redshift, the precision obtained, $\alpha(z =
1.95)/\alpha(0)~=~0.98 \pm 0.05$, corresponding to $|\alpha^{-1} d \, \alpha(t)/dt | ~<~ 7\times10^{-12}$
yr$^{-1}$, did not represent an improvement.

Since 1967, there have been many important studies of the cosmic time-dependence of $\alpha$ using quasar
absorption lines. Some of the most comprehensive and detailed investigations are described in papers by
Wolfe, Brown, \& Roberts (1976),\nocite{wolfe} Levshakov (1994),\nocite{levshakov} Potekhin \&
Varshalovich (1994), \nocite{potekhin} Cowie \& Songaila (1995),\nocite{cowie95} and Ivanchik, Potekhin,
\& Varshalovich (1999).\nocite{ivanchik} The results reported in all of these papers are consistent with
a fine-structure constant that does not vary with cosmological epoch.

 Recently, the subject has become of great interest for both
physicists and astronomers because of the suggestion that a
significant time-dependence has been found using
absorption lines from many different multiplets
 in different ions,
the `Many-Multiplet' method (see, e.g., the important papers by Dzuba, Flambaum, \& Webb 1999a,b; Webb et
al. 1999; Murphy et al. 2001a,b; and Webb et al.
2002).\nocite{dzuba99a,dzuba99b,webb01,murphy01a,murphy01b} The suggestion that $\alpha$ may be time
dependent is particularly of interest to physicists in connection with the possibility that
time-dependent coupling constants may be related to large extra dimensions (see Marciano
1984).\nocite{marciano} Unlike the previous emission line or absorption line studies, the Many-Multiplet
collaboration uses absolute laboratory wavelengths rather than the ratio of fine-structure splittings to
average wavelengths. We shall compare and contrast in \S~\ref{sec:comparison} the emission line technique
used here with the Many-Multiplet method.

\subsection{Absorption versus emission lines}
\label{subsec:absorptionversusemission}

 Why have absorption line studies of the time-dependence of
$\alpha$ dominated over emission line studies in the last three and a half years? There are two principal
reasons that absorption lines have been preferred. First, the resonant atomic absorption lines can be
observed at large redshifts since their rest wavelengths fall in the ultraviolet. By contrast, emission
line studies are limited to smaller redshifts ($z< 1.5$) since the most useful emission lines are in the
visual region of the spectrum. Second, there are many absorption lines, often hundreds, observed in the
spectra of large redshift quasars and galaxies. These lines are produced by gas clouds at many different
redshifts. There is only one set of emission lines with one redshift, representing the cosmic distance of
the source from us, for each quasar or galaxy.

Why have we chosen to return to quasar emission lines as a tool for studying the time dependence of
$\alpha$?   We shall show that emission lines can be used to make precision measurements of the time
dependence of $\alpha$ while avoiding some of the complications and systematic uncertainties that affect
absorption line measurements. In addition, the large sample of distant objects made available by modern
redshift surveys like the SDSS survey (Stoughton et al. 2002; Schneider et al. 2002)\nocite{schneider,stoughton}
and the 2dF survey
(Boyle et al. 2000; Croom et al. 2001)\nocite{boyle,croom}  provide large samples of high signal-to-noise spectra of distant
objects.

In this paper, we select algorithmically, from among 3814 quasar spectra in the SDSS Early Data Release
(EDR, Schneider et al. 2002), the quasar spectra containing O~{\sc iii}  emission lines that are most suitable
for precision measurements. A total of 95 quasar spectra pass at least
3 out of the 4 selection tests we impose on our standard sample in \S~\ref{sec:tests}. Our Standard Sample contains spectra of 42
quasars.  We also study alternative samples obtained using variants of
our standard selection algorithm. These alternative samples contain between 28 and 70 individual quasar spectra.

The techniques described in this paper could be applied to similar
transitions in [Ne~{\sc iii}] ($\lambda
\lambda ~3968, 3869$) and in [Ne~{\sc v}] ($\lambda \lambda ~3426,
3346$).  However, the [O~{\sc iii}] emission lines
are typically  an order of magnitude stronger in quasar spectra  than the just-mentioned transitions of
[Ne~{\sc iii}] and [Ne~{\sc v}] (cf. Vanden Berk et al 2001; Schneider et al. 2002). Moreover, the intensity ratio of
the [O~{\sc iii}] lines is a convenient value (about 3.0) and the relevant region of the spectrum is relatively
free of other possibly contaminating emission or absorption
lines. After [O~{\sc iii}], the [Ne~{\sc iii}] lines are
the most promising pair of emission lines for studies of the time dependence of $\alpha$.

\subsection{Outline of the paper}
\label{subsec:outlineofpaper}

 We present in \S~\ref{sec:outline} a summary of the main ingredients
in the analysis we perform. Then in \S~\ref{sec:blind} we describe how we developed the selection
algorithm and performed the analysis blindly, i.e., without knowing what the results meant
quantitatively until all of the measurements and calculations were complete. We present in
\S~\ref{sec:matching} the key procedure which we have used to  measure by how many Angstroms we have
to shift the observed 4959~\AA\ line profile to obtain a best-fit match to the 5007~\AA\ line profile.

The Standard Sample of quasar spectra that we measure is selected objectively by applying four computer
tests designed to ensure that emission line pairs that are used in the analysis will yield accurate and
unbiased estimates of $\alpha$. These selection tests are described in \S~\ref{sec:tests}.
Figure~\ref{fig:goodspectrum} illustrates a good quasar spectrum that easily passes all of the computer
tests. Figure~\ref{fig:failed} shows four spectra, each of which illustrates a failure to satisfy one of
the four standard tests.

Table~\ref{tab:average} and Table~\ref{tab:linearfit}, together with Figure~\ref{fig:standardalphavst},
present our principal results\footnote{The  value of $\left<\alpha(t)\right>$ and related quantities that
are reported in this paper differ slightly from the values reported in the version of the paper
originally submitted to the ApJ. In the original analysis, the code FitAlpha discussed in
\S~\ref{sec:matching} was incorrectly linked to SDSS data pipeline code that was temporarily under
development. In effect, one can regard this mishap with the temporary pipeline code as an additional, but
inadvertent, blind test, cf. \S~\ref{sec:blind}.} The results for our Standard Sample are explained in
\S~\ref{sec:standard}, where we discuss the average value of $\alpha$ and the best-fit slope,
$d\alpha(t)/dt$, for our Standard Sample of 42 quasars. We also describe in this section the bootstrap
method that we have used to calculate the uncertainties for the sample characteristics. In
\S~\ref{sec:differentsamples}, we describe and discuss the analysis of 17 additional samples chosen by
alternative selection algorithms. We describe the `Many-Multiplet' method in \S~\ref{sec:comparison} and
contrast the O~{\sc iii} method with the Many-Multiplet method. We summarize and discuss our results for
the SDSS Early Data Release Sample in \S~\ref{sec:discussion}.

{\bf Note added in proof.} Since this paper was written, many more quasar spectra have become available
from the SDSS Data Release One sample (see Schneider et al. 2003)\nocite{DR1}. We have analyzed this much
larger sample of quasars using exactly the same technique as described in the main body of the present
paper. We present our results for the SDSS Data Release One sample in \hbox{Appendix C}.

\subsection{How should this paper be read?}
\label{subsec:howread}

This paper contains many different things, including (not necessarily in this order): 1) a detailed
discussion of the algorithmic basis for selecting the 18 different samples that we analyze, 2) a
description of how our algorithmic measurements are made in practice, 3) a discussion of the errors in
the measurements, 4) tabulated details of the measurements sufficient to allow the reader to change or
check the data analysis,  5) quantitative results on the time dependence of the fine structure constant
in our Standard Sample, 6) quantitative results on the time dependence of $\alpha$ for 17 alternative
samples, 7) a comparison of the Many-Multiplet and O~{\sc iii} methods for studying the time dependence of the
fine structure, and 8) a summary and discussion of our main conclusions. In addition, we present an
accurate measurement of the ratio of the two forbidden transition probabilities that lead to the 5007~\AA
\ and 4959~\AA\ lines.

Most readers will only want to sample part of this fare. Here is our suggestion as to how to get the most
out of the paper in the shortest amount of time. We assume that you have read the introduction. If so,
then you probably will want to first read \S~\ref{sec:outline}, which gives an outline of the
procedure we follow,  and then look at Figure~\ref{fig:oiii}, Figure~\ref{fig:goodspectrum}, and
Figure~\ref{fig:standardalphavst}. Next read the conclusions and discussion in
\S~\ref{sec:discussion}.  If you are especially interested in the
differences between the O~{\sc iii}
method and the Many-Multiplet method, then look at Table~\ref{tab:comparison} and, if you want more
details about the differences, read \S~\ref{sec:comparison}. Our principal quantitative results are
given in \S~\ref{sec:standard} and \S~\ref{sec:differentsamples}. Unless you are an expert, you
can skip everything in \S~\ref{sec:standard} and \S~\ref{sec:differentsamples} except the
simple equations that give the main numerical results:
equation~(\ref{eq:averagestandard})--equation~(\ref{eq:slopes}). Sections~\ref{sec:blind}--\ref{sec:tests} contain
the main technical discussion of this paper. These sections are essential for the expert who wants to
decide whether we have done a good job. But, if you have already made up your mind on this question, you
can skip \S\S~\ref{sec:blind}--\ref{sec:tests}.

\section{Outline of measurement procedure}
\label{sec:outline}

\begin{figure}[!t]

\centerline{\psfig{figure=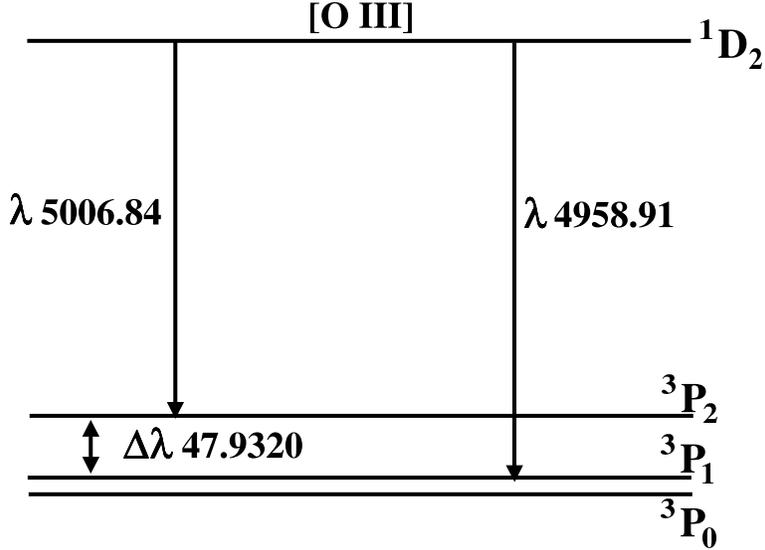,angle=270,width=4in}} \caption[]{\baselineskip=12pt Energy level
diagram for O~{\sc iii}. The figure shows the two strong emission lines that arise from the same $^1D_2$ excited
level of twice-ionized oxygen. The $5007$~\AA\  and $4959$~\AA\  lines are separated by about $48$~\AA \
by the fine-structure energies of the ground-state triplet P states. The wavelength separation $\Delta
\lambda$ divided by the sum of the wavelengths of the two emission lines is proportional to $\alpha^2$;
this dimensional-less ratio (difference divided by sum)is defined as $R$ in equation~(\ref{eq:defnR}). The
quantity $R$ can be measured in distant quasars or galaxies and is independent of cosmological epoch
unless $\alpha^2$ depends upon time. } \label{fig:oiii}
\end{figure}

 Figure~\ref{fig:oiii} shows the relevant part of the energy level diagram for
twice-ionized oxygen, O~{\sc iii}. We utilize in this paper the two
strongest nebular emission lines of [O~{\sc iii}],
$\lambda 5007$ and $\lambda 4959$, both of which originate on the same initial excited level, $^1D_2$.
Since both lines originate on the same energy level and both lines have a negligible optical depth (the
transitions are strongly forbidden), both lines have exactly the same emission line profile\footnote{The
effect of differential reddening on the line splitting can easily be shown to be $\sim 10^{-8} \tau_0$,
where $\tau_0$ is the optical depth at $5007$~\AA.  This shift is negligible, more than four orders of
magnitude less than our measurement uncertainties.}. If there are multiple clouds that contribute to the
observed emission line, then the observed emission line profiles are composed of the same mixture of
individual cloud complexes.

The two final states are separated by the fine-structure interaction. Thus the energy separation of the
$\lambda 5007$ and $\lambda 4959$ emission lines is proportional to $\alpha^4$, while the leading term in
the energy separation for each line is proportional to $\alpha^2$. To very high accuracy (see
\S~\ref{appendix:A} of the Appendix,), the difference in the wavelengths divided by the average of
the wavelengths, R,
\begin{equation}
R(t)~=~\frac{\lambda_2(t) - \lambda_1(t)}{\lambda_1(t) + \lambda_2(t)} \, , \label{eq:defnR}
\end{equation}
is proportional to $\alpha^2$. The cosmological redshift of the astronomical source that emitted the
oxygen lines cancels out of the expression for $R$, which depends only on the ratio of measured
wavelengths, either added or subtracted. Thus a measurement of $R$ is a measurement of $\alpha^2$ at the
epoch at which the [O~{\sc iii}] lines are emitted. The key element in measuring $R(t)$ is the determination of
the shift in Angstroms that produces the best-fit between the line profiles of the two emission lines.

The wavelengths in air of the nebular emission lines
\begin{equation}
\lambda_1~=~ 4958.9110~\mbox{\AA};~~~~~ \lambda_2~=~ 5006.8430~\mbox{\AA},  \label{eq:locallambdas}
\end{equation}
and their separation
\begin{equation}
\Delta \lambda ~=~ 47.9320~ \mbox{\AA}. \label{eq:localdeltalambda},
\end{equation}
have been measured accurately from a combination of laboratory measurements with a theta-pinch discharge
(Pettersson 1982)\nocite{pettersson}  and, for the wavelength separation, infrared spectroscopy with a
balloon borne Michelson interferometer (Moorwood et al. 1980).\nocite{moorwood} The conversion from
vacuum wavelengths to wavelengths in air was made using the three term formula of Peck and Reeder (1972).
\nocite{reeder} Combining equation~(\ref{eq:locallambdas}) and equation~(\ref{eq:localdeltalambda}), we have
\begin{equation}
R(0) ~=~ 4.80967\times 10^{-3}\left[1 \pm 0.00001 \right] \, ,
\label{eq:Rofzero}
\end{equation}
where the error for $R(0)$ is based primarily on measurement by Moorwood et al. (1980) of the intervals
in the ground $^3P$ multiplet (see also Pettersson 1982).\nocite{pettersson}

One can write the ratio of $\alpha^2$ at much earlier time $t$ to the value of $\alpha^2$ obtaining at
the current cosmological epoch as
\begin{equation}
\frac{R(t)}{R(0)} ~=~ \frac{\alpha^2(t)}{\alpha^2(0)} ~=~ \left[\frac{\Delta \lambda(t)}{\lambda_1(t) +
\lambda_2(t)}\right] \left[ \frac{\lambda_1(0) + \lambda_2(0)}{\Delta \lambda(0)}\right]
.\label{eq:ratioofalphas}
\end{equation}
The light we observe today was emitted by distant quasars (or galaxies) at an epoch that was $10^9$ yr or
$10^{10}$ yr earlier. This huge time difference makes possible
 precise measurements of the time dependence of $\alpha^2$ by recording the redshifted
wavelengths of the [O~{\sc iii}] emission lines emitted at a much earlier epoch.

We will compare values of R, i.e., $\alpha^2$, measured at different cosmological epochs. Although the
data are consistent with a time-independent value for $\alpha^2$, we will find an upper limit to the
possible time-dependence by fitting the measured values of $R(t)$ to a linear function of time,
\begin{equation}
R(t)~=~ R(0)[1 + S t H_0]. \label{eq:defnlinear}
\end{equation}
Here $R(t)$ is a function of $z$, $R(t(z))$, where $z$ is the cosmological redshift [see
eq.~(\ref{eq:defntime})] below for the relation between $t$ and $z$). The slope $S$ is twice the
logarithmic derivative of $\alpha^2(t)$ with respect to time in units of one over the Hubble constant.
\begin{equation}
S~=~\frac{1}{H_0\alpha^2}\left(\frac{d\alpha^2}{dt}\right)~=~\frac{2}{H_0\alpha}\left(\frac{d\alpha}{dt}\right)
\label{eq:defnslope}
\end{equation}
As defined in equation~(\ref{eq:defnlinear}) and equation~(\ref{eq:defnslope}), the slope S is independent of
$H_0$. Each redshift produces a unique value of $H_0t$.

 We do not need to know $\alpha^2(0)$, or equivalently, $\lambda_1$ and $\lambda_2$ (or $R(0))$, in order to test for the time dependence
of $\alpha$. We can use equation~(\ref{eq:ratioofalphas}) and equation~(\ref{eq:defnlinear}) to explore the time
dependence by simply taking taking $\alpha^2(0)$ to be any convenient constant. In fact, this is exactly
what we have done, which enabled us to carry out a blind analysis (see discussion in
\S~\ref{sec:blind}). For 16 of the 17 samples we have analyzed, we have not made use of the known
local value for $R(0)$, i.e., $\alpha^2(0)$ (see \S~\ref{sec:differentsamples}).

The cosmological time can be evaluated from the well known expression (e.g., Carroll and Press 1992)\nocite{carroll}
\begin{equation}
t=\frac{t_0-t_1}{H_0}=\int_0^z\frac{dz}{(1+z){\sqrt{(1+z)(1+\Omega_mz)-z(2+z)\Omega_\Lambda}}}\, ,
\label{eq:defntime}
\end{equation}
where $\Omega_m$ and $\Omega_\Lambda$ are the usual present-epoch fractional contributions of matter and
a cosmological constant to the cosmological expansion. The translation between the Hubble constant,
$H_0$, and a year is given by
\begin{equation}
\frac{1}{H_0}~=~ 1.358 \times 10^{10} {\rm ~yr^{-1}} \left(\frac{72~{\rm km
\,s^{-1}Mpc^{-1}}}{H_0}\right). \label{eq:h0}
\end{equation}
In what follows, we shall adopt the  value for $H_0$ determined by the HST Key Project, i.e., $H_0 =
72~{\rm km \, s^{-1}Mpc^{-1}}$ (Freedman et al. 2001).\nocite{freedman} If the reader prefers a different
value for $H_0$, all times given in this paper can be rescaled using equation~(\ref{eq:h0}). Except where
explicitly stated otherwise in the remainder of the paper, the time $t$ is calculated from
equation~(\ref{eq:defntime}) and equation~(\ref{eq:h0}) for a universe with the present composition of
$\Omega_m=0.3,\,\, \Omega_\Lambda=0.7$.

\section{Blind analysis}
\label{sec:blind}

To avoid biases, we performed a blind analysis.

Measurements of $\alpha^2$ were computed relative to an artificial value of $\alpha^2(0)_{\rm
artificial}=1.0 $. We did not renormalize the measured values of $\alpha^2(t)$ until the analysis was
complete.  In fact, we did not search out the precise values of the wavelengths given in
equation~(\ref{eq:locallambdas}) until after we had finished all our calculations of $\alpha^2(t)$ from the
measured wavelengths. Throughout all the stages when we were determining the selection criteria for the
sample described in \S~\ref{sec:tests}, we worked with measured values for $\alpha^2(t)$ that were
about 0.0048 [cf. eq.~(\ref{eq:Rofzero})]. At this stage, we had not researched the precise values of the
local wavelengths for the [O~{\sc iii}] emission lines. Therefore, we did not know whether the measured values
that we were getting for $\alpha^2(t)$ were close to the local value and, if so, how close.

To further separate the measuring process from any knowledge of the local value for $\alpha(0)$, we
developed the criteria for selecting the sample using a subset of the quasars in the SDSS EDR sample. We
used  in this initial sample 313 quasars with redshifts $z < 0.80$. After finalizing the selection
criteria, we applied the criteria to the remaining 389 quasars with $z < 0.80$ in the EDR. The acceptance
rate for the initial sample was 23 objects out of 313 or $7.7$\%, while the acceptance rate for the
second sample was 19 out of 389 or $5.1$\%.  We will refer later to the combined sample of 42 objects
selected in this way as our `standard sample.'

The estimates of $\alpha^2$ derived for the two parts of the standard sample were in excellent agreement
with each other. For the initial sample of 23 objects, we found

\begin{equation}
\left<\alpha^2\right> ~=~1.0002 \pm 0.0004 \, ,\label{eq:initialaverage}
\end{equation}
and for the second sample of 19 objects we found

\begin{equation}
\left<\alpha^2\right> ~=~1.0001 \pm 0.0004 \, .
 \label{eq:secondaverage}
\end{equation}
The sample average for the 42 objects is $1.0001 \pm 0.0003$ (cf. row one of  Table~\ref{tab:average} in
\S~\ref{sec:standard}).

\section{Matching the 5007~\AA\ and 4959~\AA\ line shapes}
\label{sec:matching}

The key element in our measurement of $\alpha$ is the procedure by which we match the shapes of the
5007~\AA\ and 4959~\AA\ emission line profiles. What we want to know is by how many Angstroms do we have
to shift the 4959~\AA\ line so that it best fits the measured profile of the 5007~\AA\ line. In the
process, we also determine a best-estimate for the ratio of the intensities in the two lines.

Figure~\ref{fig:oiii} shows that both of the emission lines, [O~{\sc iii}] 5007~\AA\ and 4959~\AA,  arise
from the same initial atomic state. Hence both lines have (within the accuracy of the noise in the
measurements) the same line profile, i.e., the same shape of the curve showing the flux intensity versus
wavelength, although the two lines are displaced in wavelength and have different intensities.

We have developed a computer code, FitAlpha, that finds the best match of the shapes of the 5007~\AA \
and 4959~\AA\ lines in terms of two parameters, $\eta$ and $A$. The two parameters describe how much the
measured profile of the 4959~\AA\ line must be shifted in wavelength, $\eta$,  and how much the
4959~\AA\ profile must be increased in amplitude, $A$, in order to provide the best possible
match to the profile of the 5007~\AA\ emission line.

We first discuss in \S~\ref{subsec:amplitudes} the measurements of the amplitude ratios $A$ and then
describe in \S~\ref{subsec:etaandR} the measurements of the quantity that describes the wavelength shift,
$\eta$. We show in \S~\ref{subsec:correlation} that the measured values of $A$ and $\eta$ are not
significantly correlated. We conclude by describing in \S~\ref{subsec:individualerrors} the errors that
we assign to individual measurements of $\alpha^2$.

\subsection{Amplitude ratios A} \label{subsec:amplitudes}

In the limit in which the profiles of the 5007~\AA\ and 4959~\AA\ lines  match identically and there is
no measurement noise, the quantity $A$ is the ratio of the area under the 5007~\AA\ line to the area
under the 4959~\AA\ line. More explicitly, in the limit of an ideal match of the two line profiles,
\begin{equation}
A ~=~ \frac{\int_{5007} \left[ {\rm LineFlux(\lambda) - ContinuumFlux(\lambda) } \right] d\lambda }
{\int_{4959} \left[ {\rm LineFlux(\lambda) - ContinuumFlux(\lambda) } \right] d\lambda }\, .
\label{eq:defnA}
\end{equation}
 In practice, the best-fit ratio of the amplitudes, which is measured
 by FitAlpha,  is better determined than the area under the emission lines.
 Thus, the values we report for $A$ (e.g. in Table~\ref{tab:standard} in \S~\ref{appendix:B} of the Appendix) are the best-fit ratios of the amplitudes.
  Although $A$ is one of the outputs of FitAlpha, we do not make use of $A$ except as a {\it post facto} sanity check
that the fits are sensible.

\begin{figure}[!t]
\centerline{\psfig{figure=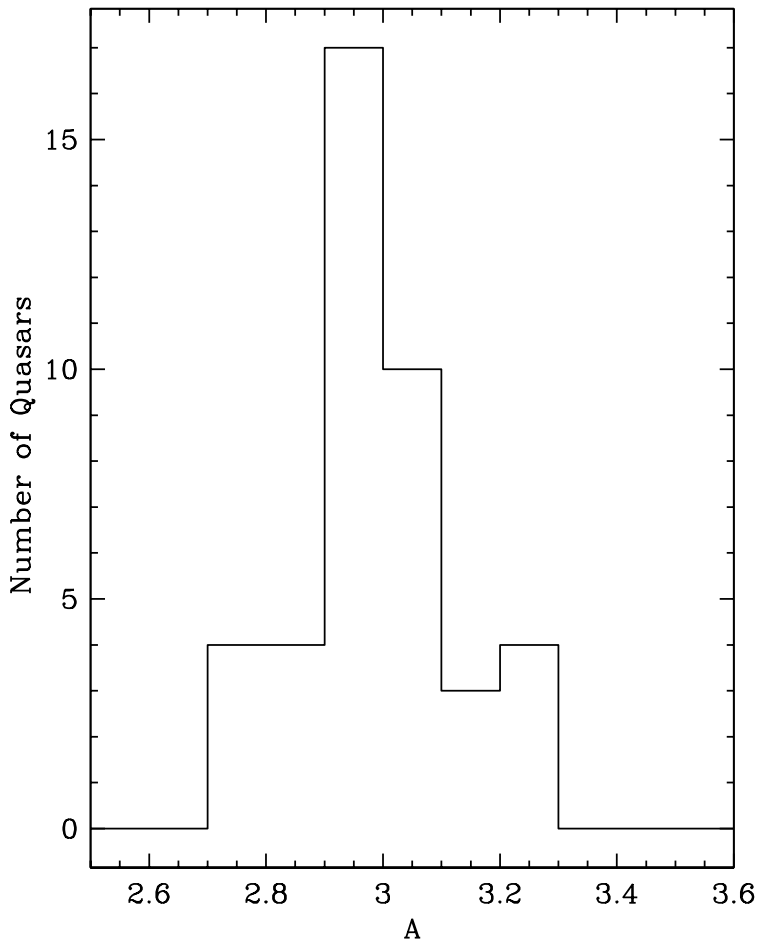,width=4in}} \caption[]{\baselineskip=12pt The ratio of amplitudes,
$A$. The figure shows a histogram for our standard sample of quasars of the ratio $A$, which is the
best-fit ratio of the amplitude of the 5007~\AA\ profile to the amplitude of the 4959~\AA\ profile. The
measured average value of $A$ is in good agreement with the expectation based upon the tabulated data in
the NIST Atomic Spectra Database [see eq.~(\ref{eq:Ameasurement}) and eq.~(\ref{eq:einsteinratio}) and
related text]. \label{fig:A} }
\end{figure}

Figure~\ref{fig:A} presents a histogram of the values of A that were measured for quasars in our standard
sample, which is defined in \S~\ref{sec:tests}. Our measurements have a typical $1\sigma$ spread of
about $\pm 0.1$.  The sample average of the measurements for A, and the bootstrap uncertainty determined
as described in \S~\ref{subsec:bootstrap}, are

\begin{equation}
\left<A_{\rm measured}\right> ~=~ 2.99 \pm 0.02 \, . \label{eq:Ameasurement}
\end{equation}

The quantity $\left<A_{\rm measured}\right>$ is equal to the ratio of the decay rates of the 5007~\AA \
and the 4959~\AA\ lines, i.e., the ratio of the Einstein A coefficients. Thus the measured value  for
the ratio of the transition probabilities corresponding to the 5007~\AA\ and 4959~\AA\ lines is

\begin{equation}
\frac{A\left(5007\right)}{A\left(4959\right)} ~=~ 2.99 \pm 0.02\,. \label{eq:einsteinratio}
\end{equation}

The measured value of ${A\left(5007\right)}/{A\left(4959\right)}$ is consistent with the best theoretical
estimate given in the NIST Atomic Spectra Database, ${A\left(5007\right)}/{A\left(4959\right)} = 2.92$
(Wiese, Fuhr, and Deter 1996)\nocite{wiese}. The quoted NIST accuracy level for this theoretical estimate is `B', which
generally corresponds to a numerical accuracy of better than 10\%. Our measurement is more accurate than
the theoretical estimate.

\subsection{Measuring $\eta$ and R}
\label{subsec:etaandR}

 We define $\eta$ by the relation
\begin{equation}
\eta ~\equiv~ \frac{\lambda_2}{\lambda_1} -1 ,
 \label{eq:defneta}
\end{equation}
where $\lambda_1$ and $\lambda_2$ are the measured values of the redshifted emission lines corresponding
to the lines with rest wavelengths given by equation~(\ref{eq:locallambdas}). We use the measured values of
$\eta$ to calculate the value of $\alpha^2(z)$ for each source. The ratio R defined by
equation~(\ref{eq:defnR}) can be written in terms of $\eta$ as

\begin{equation}
R ~=~ \frac{\eta}{\eta + 2}. \label{eq:rintermsofeta}
\end{equation}

FitAlpha finds the best-fit values of $\eta$ as follows. The Princeton 1-D reduction code (Schlegel
2003\nocite{schlegel}) determines the redshift of the quasar by fitting the entire quasar spectrum to a
template. This fit yields an approximate value for the redshift that locates the 4959~\AA\ and 5007~\AA \
lines to within 2-3 spectral pixels.  Each of the spline-resampled pixels has a width of $69.0 {\rm \,
km\,s^{-1}}$. FitAlpha selects a 20 pixel wide region of the spectrum centered around the expected center
of the 4959~\AA\ line and a wider (35 pixel) region around the expected center of the 5007~\AA\ line. The
Spectro-2D code represents each quasar spectrum as a $4^{\rm th}$ order B-spline, where the spline points
are spaced linearly in log-wavelength (Burles \& Schlegel 2003\nocite{burlesschlegel}), $\log \lambda_1
= {\rm const.} + 10^{-4}\times i$. The optical output of the SDSS spectrograph gives pixel elements that
are very nearly proportional to $\log \lambda$.  This B-spline can be evaluated at any choice of
wavelengths. We chose to evaluate each spectrum on a dense grid spaced in units of $10^{-5}$ in
log-wavelength ($\sim 0.1$ pixel), corresponding to approximately 6.9${\rm ~km\, s^{-1}}$. The measured
average full width at half maximum of the 5007~\AA\ lines in our standard sample is 5.5 pixels ($\sim
9$~\AA).

The SDSS spectra cover the wavelength region 3800--9200 \AA \,. The resolution (FWHM) varies from $140
{~\rm to~} 170 \rm \, km\,s^{-1}$, with a pixel scale within several percent of $69{\rm \, km\,s^{-1}}$
everywhere (Burles \& Schlegel 2003). The wavelength solution is performed as a simultaneous fit to an
arc spectrum observation and the sky lines as measured in each object observation.  The arc lines
constrain the high-order terms in the wavelength calibration, while the sky lines constrain any small
flexure terms and plate scale changes between the arc and object observations.  The RMS in the recovered
arc and sky line positions is measured to be approximately $1 {\rm \, km\,s^{-1}}$.  Therefore, it is
possible that systematic errors could remain in the wavelength
calibration at this level
(Burles \&
Schlegel 2003). The flux-calibration has been shown to be accurate to a few percent on average, which is
impressive for a fiber-fed spectrograph (Tremonti et al. 2003\nocite{tremonti}, in preparation). The
remaining (small) flux-calibration residuals are coherent on scales of 500 \AA \,, which has negligible
effect on our line centers which are measured using regions $< 100$ \AA \ wide.

The measured profile of the 4959~\AA\ line is compared with each of the spline-fit representations of
the 5007~\AA\ region.  We make the comparison only within a region of $2\times10^{-3}$ in log-wavelength
(about 23~\AA). For each possible shift between the two emission lines in log-wavelength, we minimize
over the multiplicative scale factor $A$ between the profiles of the two lines and a quadratic
representation of the local continuum. This continuum is primarily the approximately power-law of the
quasar spectrum and the wings of the H$\beta$ line.  The
 value of $\eta$ [cf. eq.~(\ref{eq:defneta})] is
determined by minimizing the $\chi^2$ value for the fit between the two line shapes.

\subsection{Are $\eta$ and A correlated?}
\label{subsec:correlation}

It is natural to ask if the measured values of $\eta$ and A are correlated since they are both determined
by the same computer program. A priori, one would not expect the two variables to be correlated if the
best shift and  the best rescaling are really disjoint. For the comparison of two lines with the same
intrinsic shape, the shift and the rescaling should be independent of each other. However, computer
programs that process real experimental data do not always yield the expected answer.  Hence, we have
checked whether there is a significant correlation between the observed values of $\eta$ and A.

\begin{figure}[!t]
\centerline{\psfig{figure=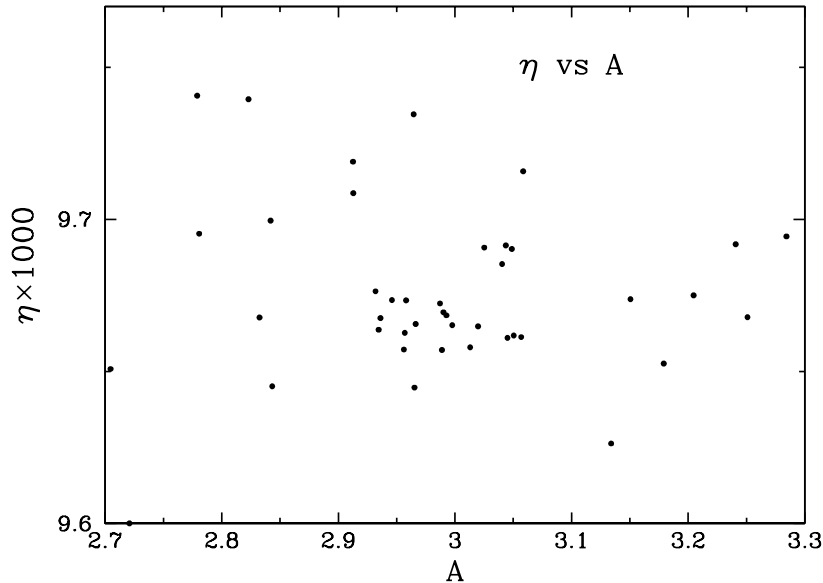,width=5in,angle=270}} \caption[]{\baselineskip=12pt Lack of
correlation between $\eta$ and A. The plotted points represent the measured values of A and
$1000\times\eta$ for all 42 quasars in our standard sample. \label{fig:alphavseta} }
\end{figure}

Figure~\ref{fig:alphavseta} shows all the measured values of (A,$\eta$) for all 42 quasars in our
standard sample. The figure looks like a scatter diagram and indeed the computed linear correlation
coefficient is only -0.034.  If there were no intrinsic correlation between $\eta$ and A, one would
expect just by chance to obtain a correlation coefficient bigger than the measured value in 83\% of the
cases. We conclude that there is no significant correlation between $\eta$ and A.

\subsection{Error estimates on individual measurements}
\label{subsec:individualerrors}

 The errors on the measured SDSS fluxes are not perfectly correct; the
pipeline software overestimates the errors for lower flux levels. We correct the errors approximately by
multiplying all of the  errors by a constant factor (typically of order 0.75) chosen to make $\chi^2$ per
degree of freedom equal to $1$ for the best-fit value of $\eta$. These adjusted  errors are used to
evaluate the uncertainty in the measured value of $\eta$. However, we show in \S~\ref{sec:standard}
and in \S~\ref{subsubsec:errors} that the bootstrap error for the sample average value of $\alpha$
and of the first derivative of $\alpha$ with time are very insensitive to the estimated uncertainties in
the individual measurements.

The error on the measured value of $\alpha^2$ is dominated by the error on the measurement of the
wavelength separation, $\Delta \lambda $, between the 5007~\AA\ and the 4959~\AA\ line.  Before making
the measurements, we estimated, based upon prior experience with CCD spectra, that we should be able to
measure $\Delta \lambda$ to about 0.05 of a pixel. Since the width of each pixel is about 70 ${\rm
km\,s^{-1}}$, or about 1.2~\AA\ at 5000~\AA, we expected to be able to make measurements of
$\alpha^2$ that were accurate to about $1.2\times10^{-3}$. As we shall see in \S~\ref{sec:standard}
(cf. Table~\ref{tab:standard} of \S~\ref{appendix:B} of the Appendix), the average quoted error for
an individual measurement of $\alpha^2$ is $1.8\times10^{-3}$, somewhat larger (but not much larger) than
our expected error.

\section{Sample selection tests}
\label{sec:tests}

We begin this section with some introductory remarks, made in \S~\ref{subsec:introsample}, about the
sample and the sample selection. In the last four subsections of this section,
\S\S~\ref{subsec:kstest}--\ref{subsec:signaltonoise}, we define four standard tests that were
used to select the standard sample of quasar spectra from which we have determined our best estimates of
$\alpha^2$ and $\alpha^{-2} d\alpha^2/dt$. Our standard sample, discussed in \S~\ref{sec:standard},
passes all four of the tests as described here. We present in \S~\ref{subsec:alternatives} a number
of variations of the standard tests.

\subsection{Introductory remarks about the spectra and the sample selection}
\label{subsec:introsample}

\subsubsection{Spectra used}
\label{subsubsec:spectra}

We list in Table~\ref{tab:whowhat}  of \S~\ref{appendix:B} of the Appendix each of the 95 quasars
in the SDSS EDR sample that have passed at least three of the four standard tests that we describe later
in this section. Table~\ref{tab:whowhat} shows which of the four tests each quasar satisfies. The table
also gives the measured value of $\eta$ [see eq.~(\ref{eq:defneta})] for every quasar that we have
considered. If the reader would like to perform a different data analysis using our sample, this can
easily be accomplished with the aid of Table~\ref{tab:whowhat},
eqs.~(\ref{eq:defnR})--(\ref{eq:defnslope}), and equation~(\ref{eq:rintermsofeta}).

Although the spectra that we use here were obtained from the SDSS Early Data Release, the spectral
reductions that we use are significantly improved over the reductions given in the EDR. Improvements in
the reduction code are described in Burles and Schlegel (2003)\nocite{burlesschlegel}.

As described in \S~\ref{sec:blind}, the four standard tests were developed using an initial sample of 313
quasars in the EDR data of SDSS that could potentially show both of the [O~{\sc iii}] emission lines. The
range of redshifts that was included in this initial sample is $z = 0.163$ to $z = 0.799$. The complete
EDR sample of quasars that we have studied includes 702 quasars with redshifts ranging from $z = 0.150$
to $0.799$. A total of 23 out of 313 quasars in our initial sample passed all four of the tests described
below. In our final sample, 42 quasars passed all four tests. In what follows, we shall refer to the 23
accepted quasar spectra as our ``initial sub-sample" and the 42 total accepted quasars as our ``standard"
sample.

Unfortunately, the majority of the EDR spectra that we examined were not suitable for measurement.  In
nearly all of these unsuitable spectra, the O~{\sc iii} lines did not stand out clearly above the continuum or
above the noise. In a very few cases, we encountered a technical problem processing the spectra. As the
first step in our `blind analysis', before we imposed our four standard tests, we threw out unsuitable
spectra by requiring that both of the O~{\sc iii} lines have a positive equivalent width and that the
fractional error in $\alpha^2$, determined from equation~(\ref{eq:ratioofalphas}), be less than unity. Only
260 out of the 702 quasar spectra (37\%) in which the O~{\sc iii} lines could have been measured passed this
preliminary filtering and were subjected to the standard four tests to determine the standard sample.
{\it Post facto}, we checked that our results were essentially unchanged if we omitted this preliminary
cut on unsuitable spectra, namely: the standard sample we analyze increased from 42 to 45 objects, the
sample average of $\left<\alpha(z)\right>$  decreased by 0.002\%, and the calculated error bar increased
by 0.4\%. We chose to impose the preliminary cut on suitable spectra because it increased our speed in
analyzing a variety of alternative cuts discussed in \S~\ref{sec:differentsamples}.

\subsubsection{Standard cuts}
\label{subsubsec:standards}

 The standard cuts described in the following subsections were imposed in the `blind' phase of our
analysis in order to make sure that the accepted spectra were of high quality and that the line shapes of
the 5007~\AA\ and 4959~\AA\ lines were the same.  Although we were seeking to eliminate any sources of
biases by eliminating problematic spectra before we made the measurements, we have found {\it post facto}
no evidence that omitting the any of the cuts would have led to a systematic bias.

We wanted to guard  against contamination by H$\beta$ emission and to ensure that the signal to noise
ratio was high. We also devised two tests that checked on the similarity of the two line shapes, the KS
test discussed in \S~\ref{subsec:kstest} and the line peak test described in
\S~\ref{subsec:liningup}. In practice, all of these tests, with the exception of the check on
H$\beta$ contamination, are designed to prevent noise fluctuations or unknown measuring errors from
distorting the shape of one or both of the emission line profiles.

We shall see in \S~\ref{sec:differentsamples} that we could have been less stringent in imposing
{\it a priori} criteria for acceptance of quasar spectra. We describe in
\S~\ref{sec:differentsamples} the results that were obtained for four separate samples in which we
omitted a different one of each of the  four standard tests described in
\S~\ref{subsec:kstest}--\S~\ref{subsec:signaltonoise}. We also present the results of analyses
made using a variety of weaker versions of the four selection tests discussed in the present section.
Although we obtain larger samples by using less stringent acceptance criteria, the final results for the
time dependence of $\alpha$ are robust, essentially independent of the form and number of the selection
tests. Essentially, we can trade off spectral quality versus number of accepted spectra without
significantly changing the final conclusion.

\begin{figure}[!t]
\centerline{\psfig{figure=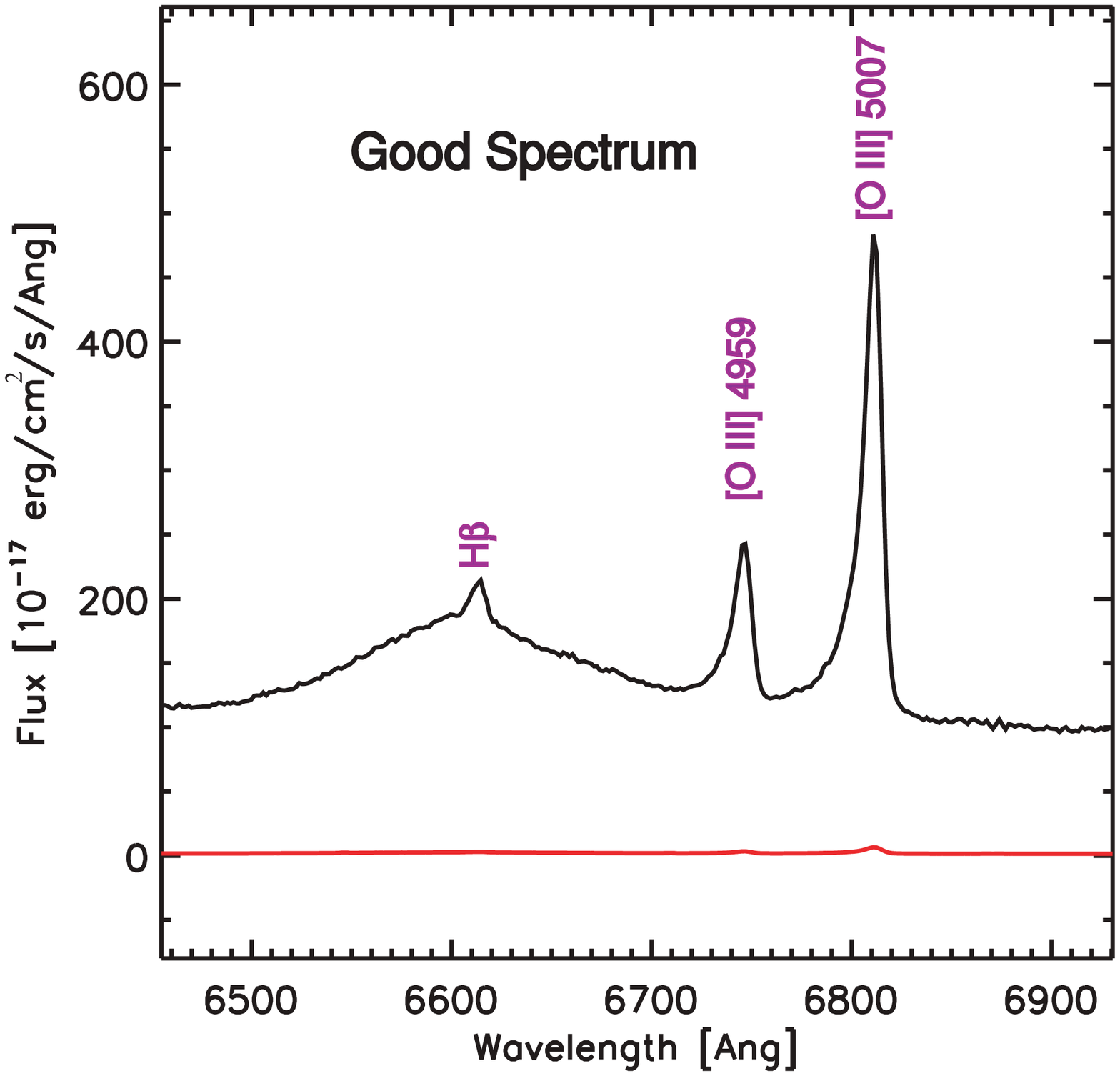,width=3in}\hglue.5in\psfig{figure=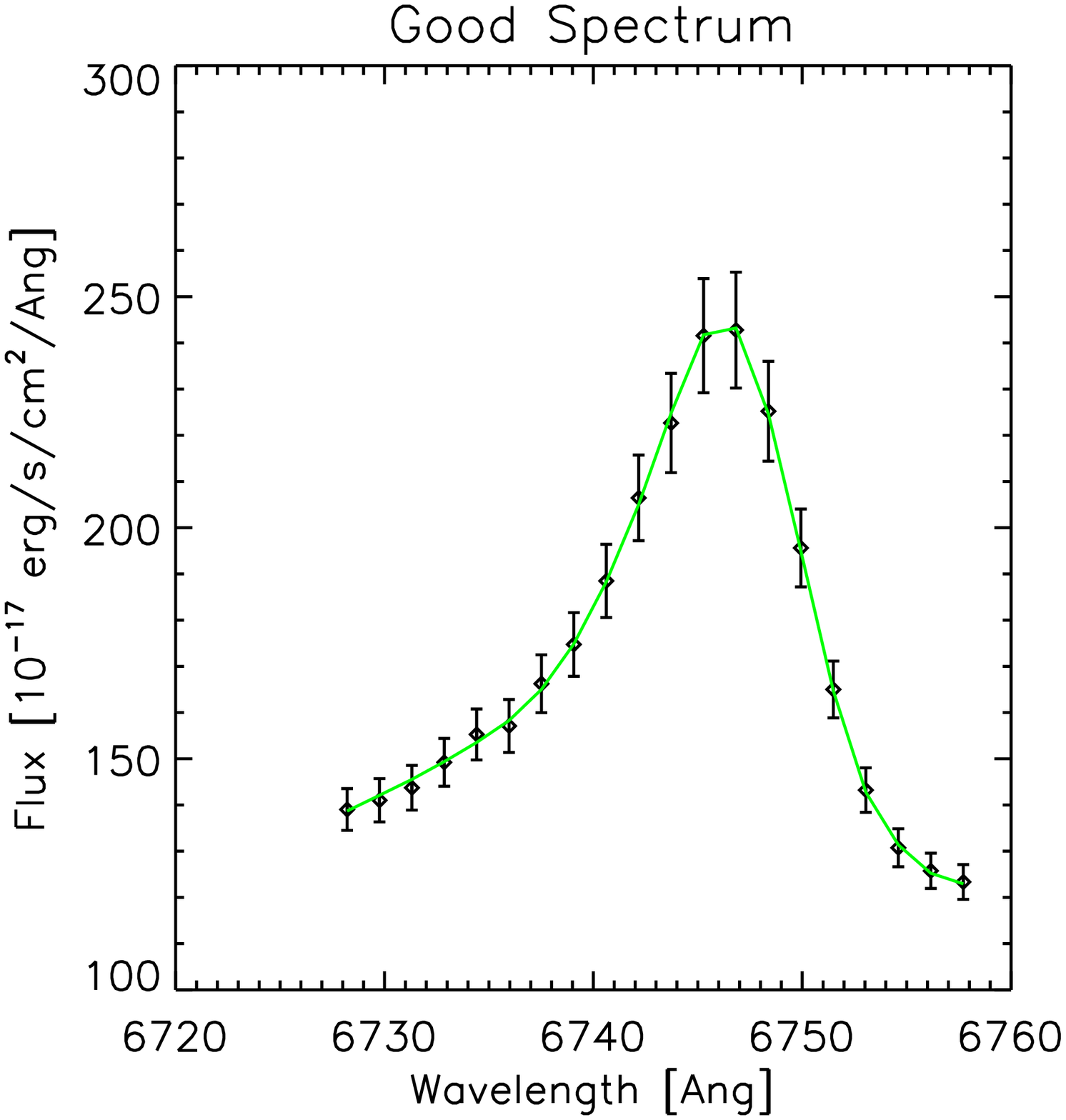,width=3in}}
\caption[]{\baselineskip=12pt A good spectrum. The figure shows the spectrum of SDSS 105151-005117, which
is also PG 1049-005. The quasar has a redshift of z = 0.359. The quasar spectrum easily passes all four
of the tests described in \S~\ref{sec:tests}.  The left panel shows the measured quasar energy
spectrum in the wavelength region of interest. The approximately horizontal line near the zero flux level
represents the estimated total error in the flux at each pixel. The right hand panel shows as a
continuous curve the best-fit spline to the measured shape of the 5007~\AA\ line, after shifting the
line center ($\eta$), decreasing the amplitude ($A$),  and super-imposing the spline fit on the measured
data points of the 4959~\AA\ line. \label{fig:goodspectrum} }
\end{figure}

 Figure~\ref{fig:goodspectrum} shows the spectrum of SDSS 105151-005117, z = 0.359,
 which easily passes all of the selection criteria described
below. If all of the quasar spectra were as clean and as well measured as the spectrum of SDSS
105151-005117, no selection tests would be required. We also show in the right hand panel of
Figure~\ref{fig:goodspectrum} the continuous spline fit of the shape of the 5007~\AA\ line that, when
the center is shifted in wavelength and the amplitude of the curve is decreased by a constant factor
($A$), best-matches the measured data points for the 4959~\AA\ line shape.

\begin{figure}[!t]
\centerline{\psfig{figure=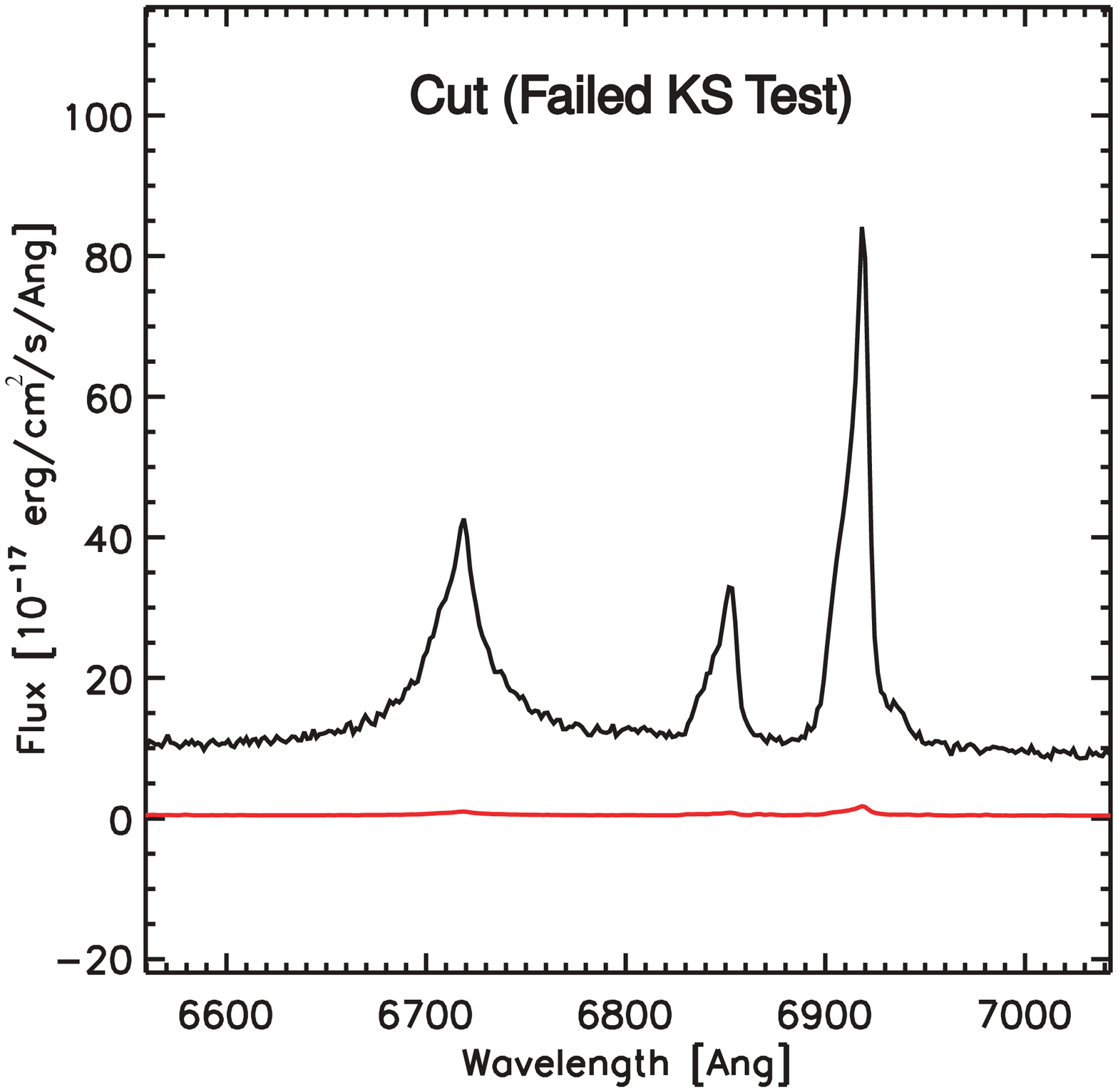,width=3in}\hglue.5in\psfig{figure=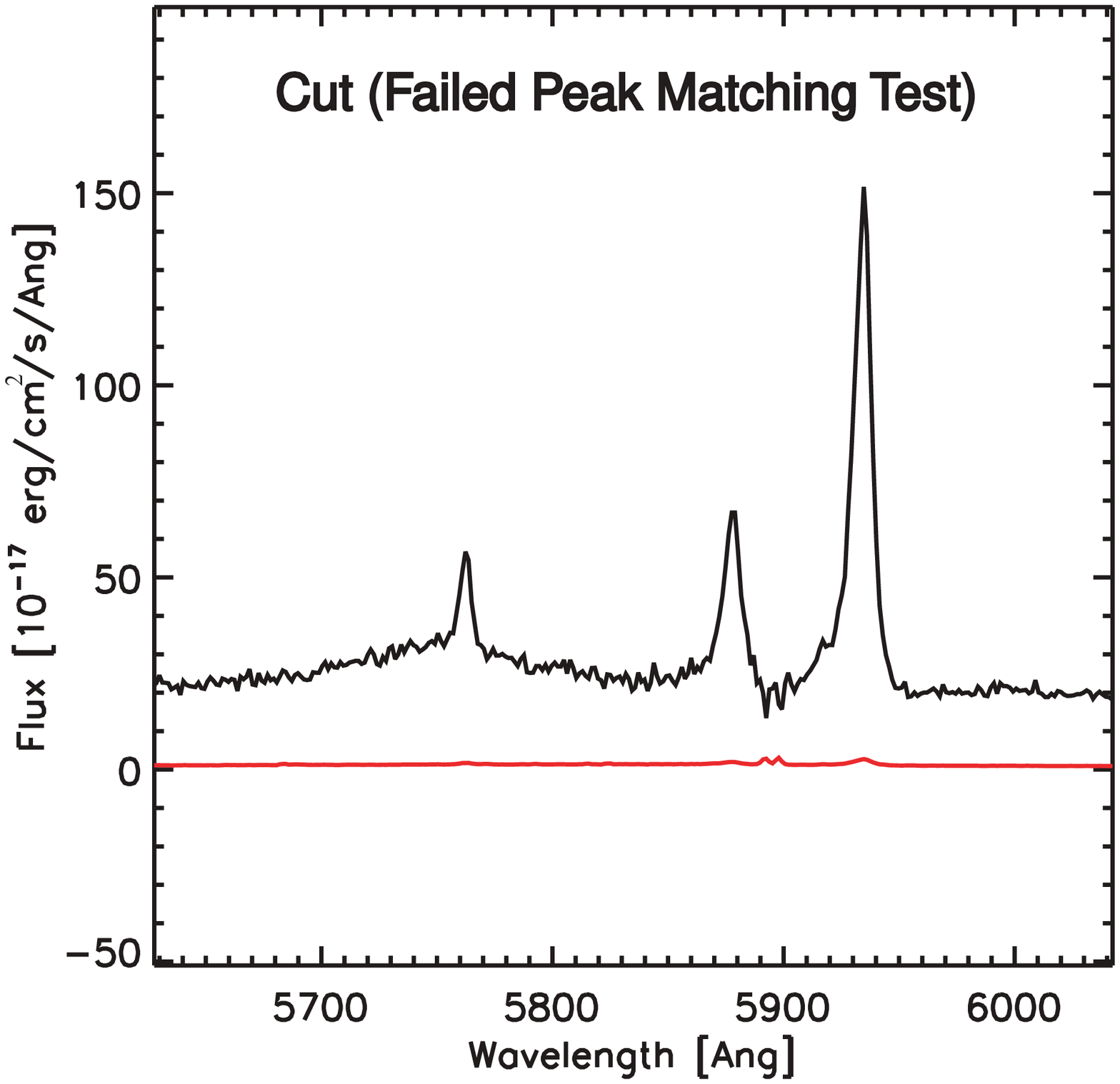,width=3in}}
\centerline{\psfig{figure=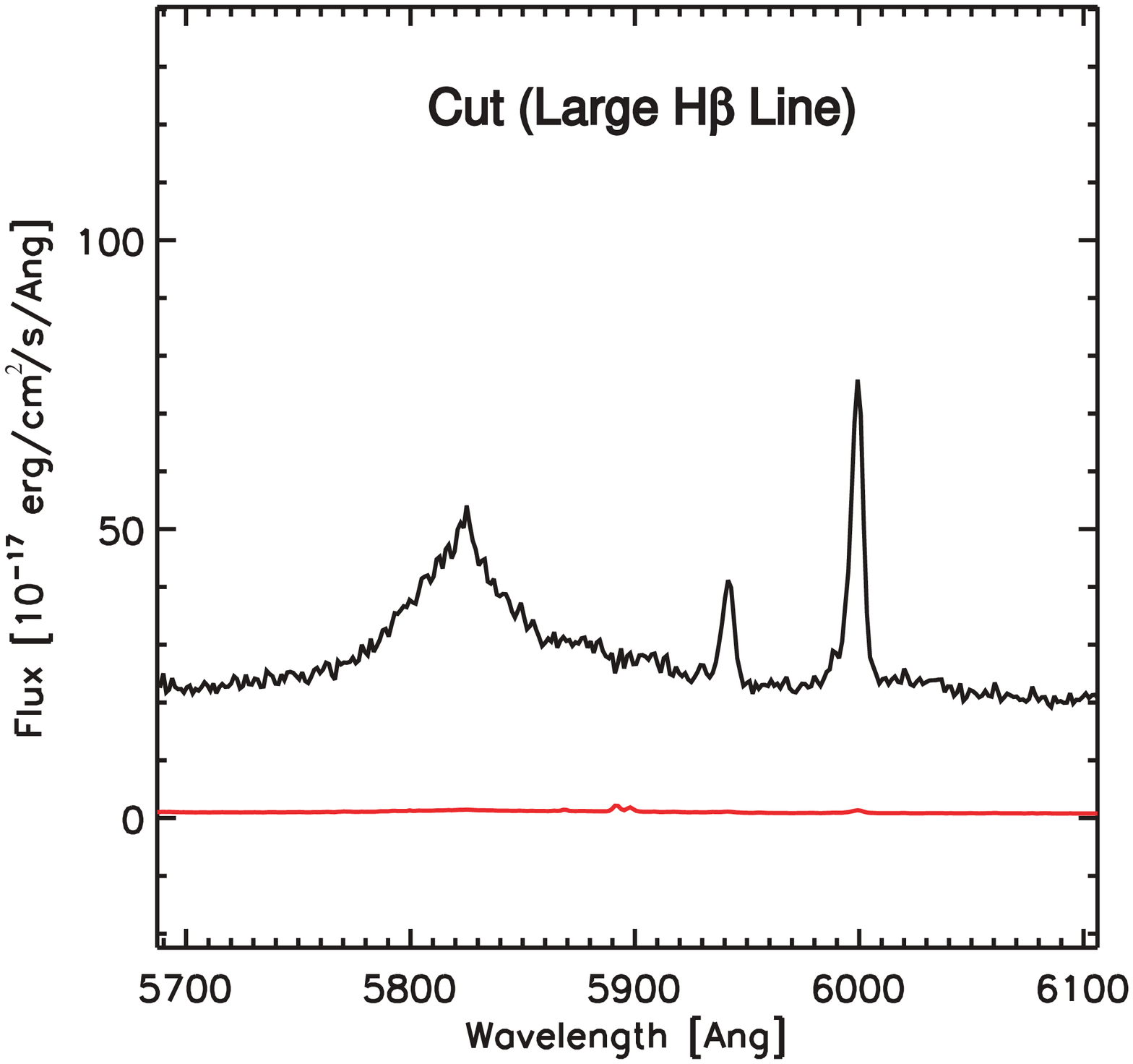,width=3in}\hglue.5in\psfig{figure=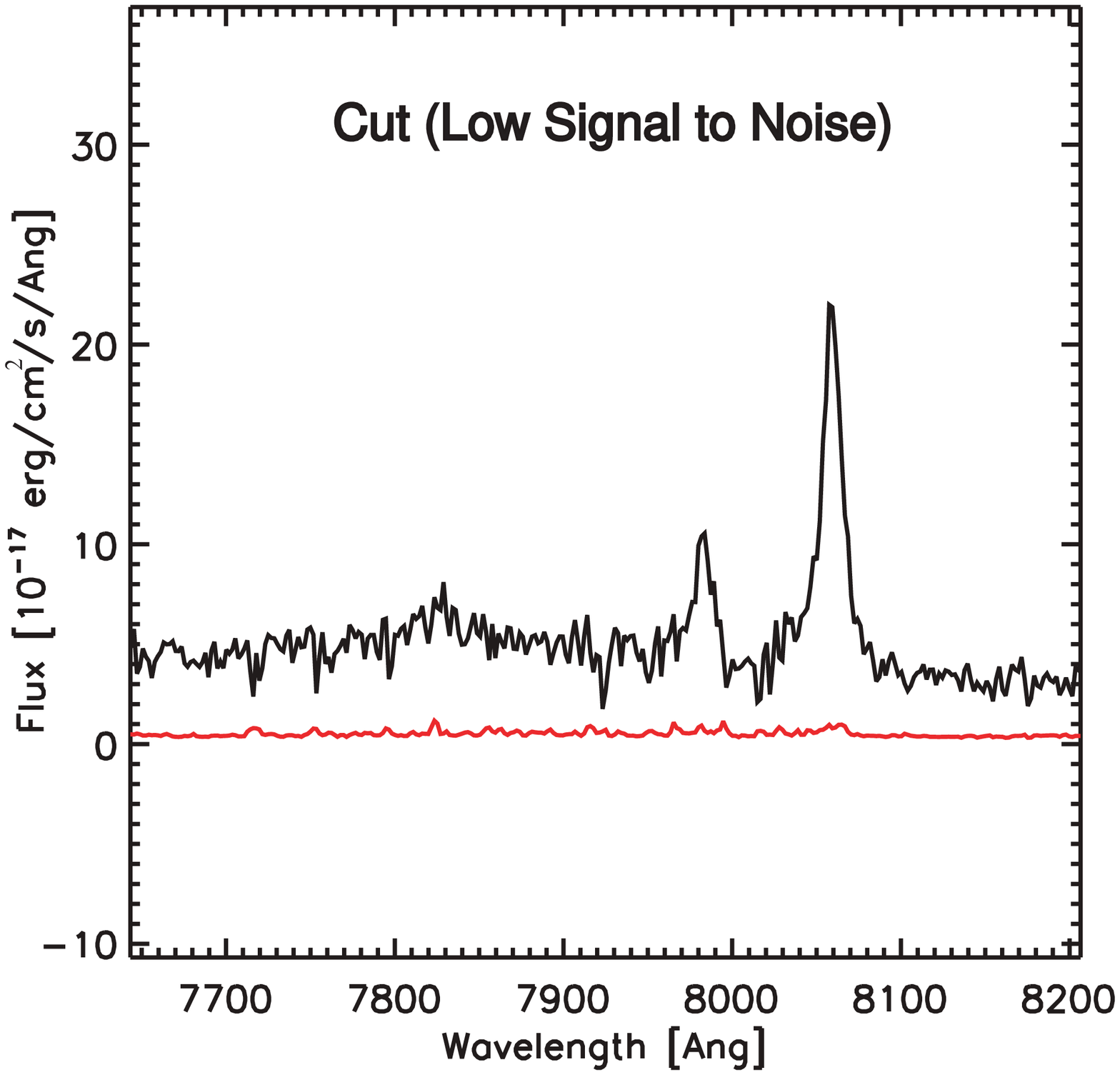,width=3in}}
\caption[]{\baselineskip=12pt Examples of four failed spectra.  The figure shows examples of spectra that
failed each of the tests described in \S~\ref{sec:tests}. The spectrum in the upper left panel
failed the KS test (cf. \S~\ref{subsec:kstest});  the upper right panel shows a spectrum in which
the peaks failed to line up correctly (cf. \S~\ref{subsec:liningup}).  The lower left panel displays
a spectrum in which the H$\beta$ line is so large that it might
contaminate the [O~{\sc iii}] 4959~\AA\ (cf.
\S~\ref{subsec:hbeta}); the lower right panel shows a spectrum in which the signal to noise ratio is
too low to permit an accurate measurement of the splitting between the
two [O~{\sc iii}] emissions lines (cf.
\S~\ref{subsec:signaltonoise}) \label{fig:failed}.}
\end{figure}

\begin{figure}[!t]
\centerline{\psfig{figure=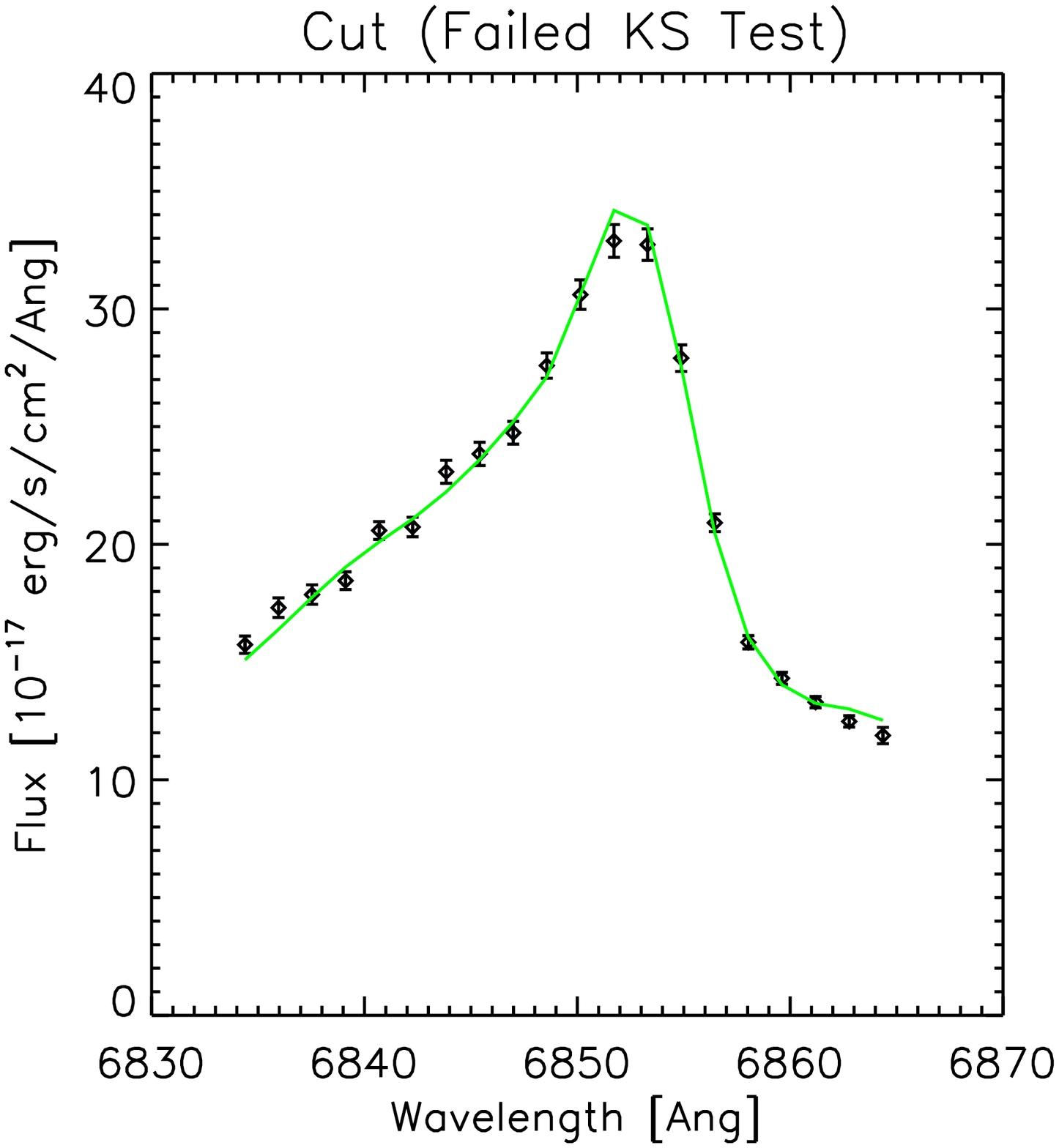,width=3in}\hglue.5in\psfig{figure=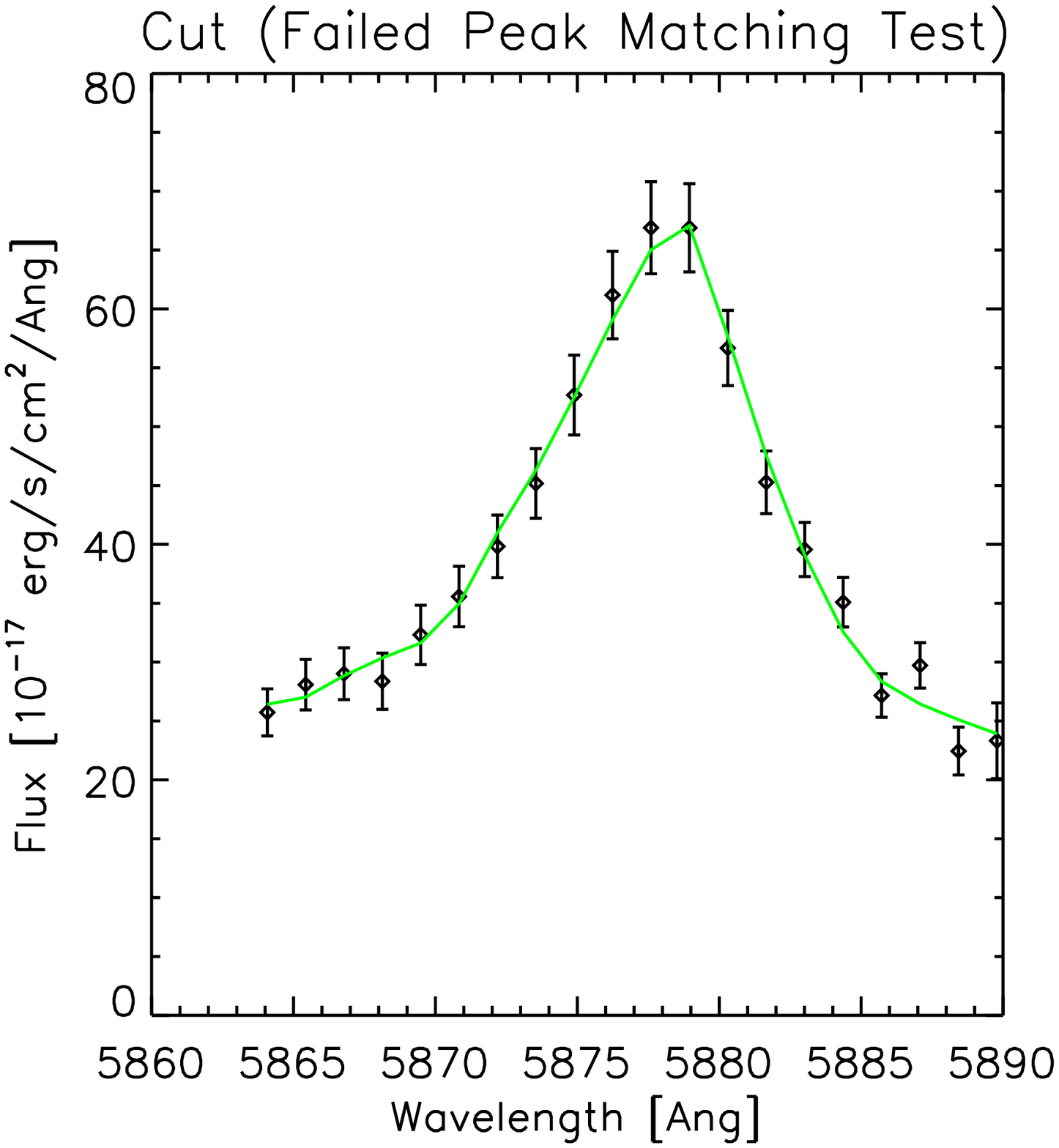,width=3in}}
\centerline{\psfig{figure=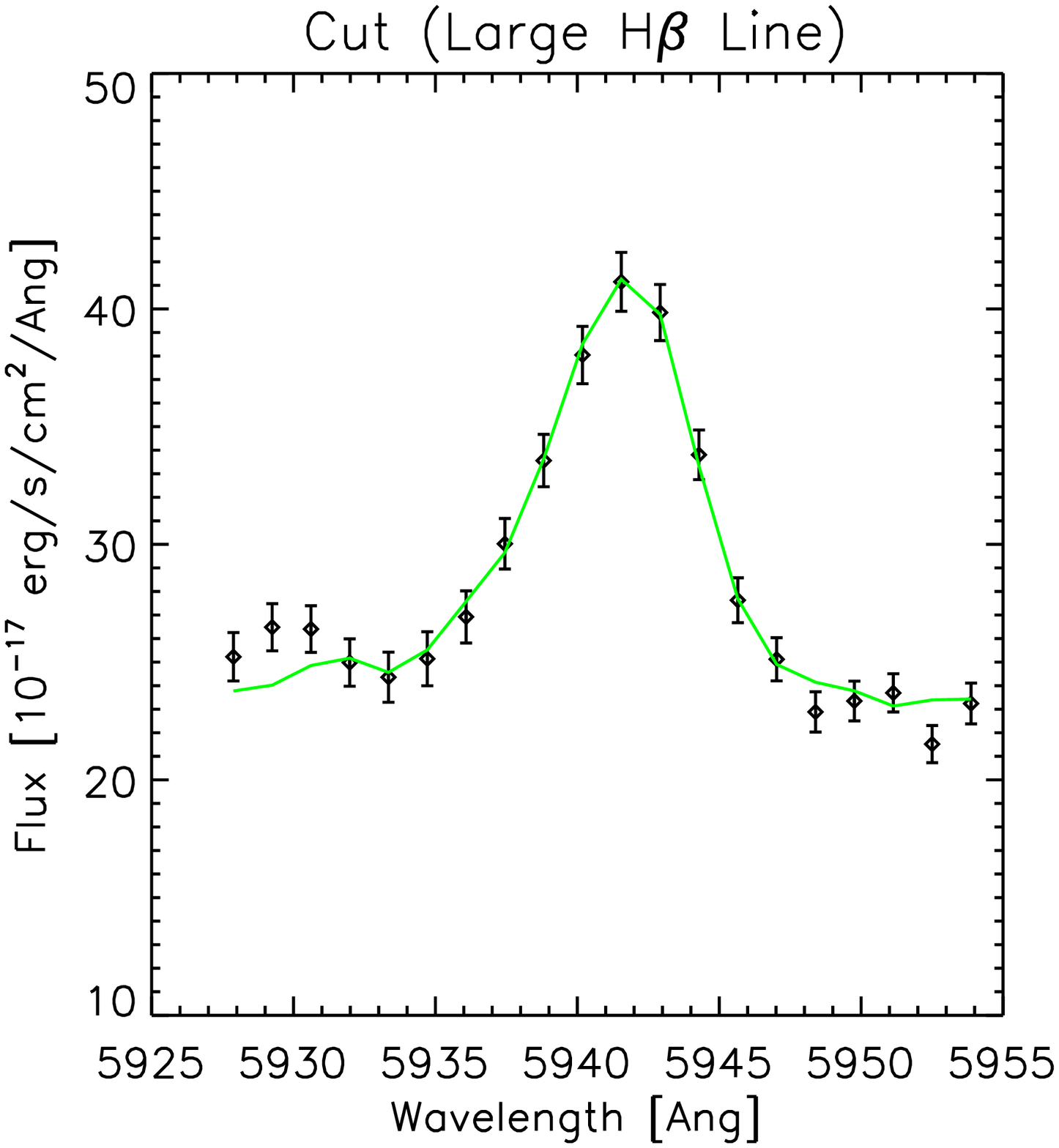,width=3in}\hglue.5in\psfig{figure=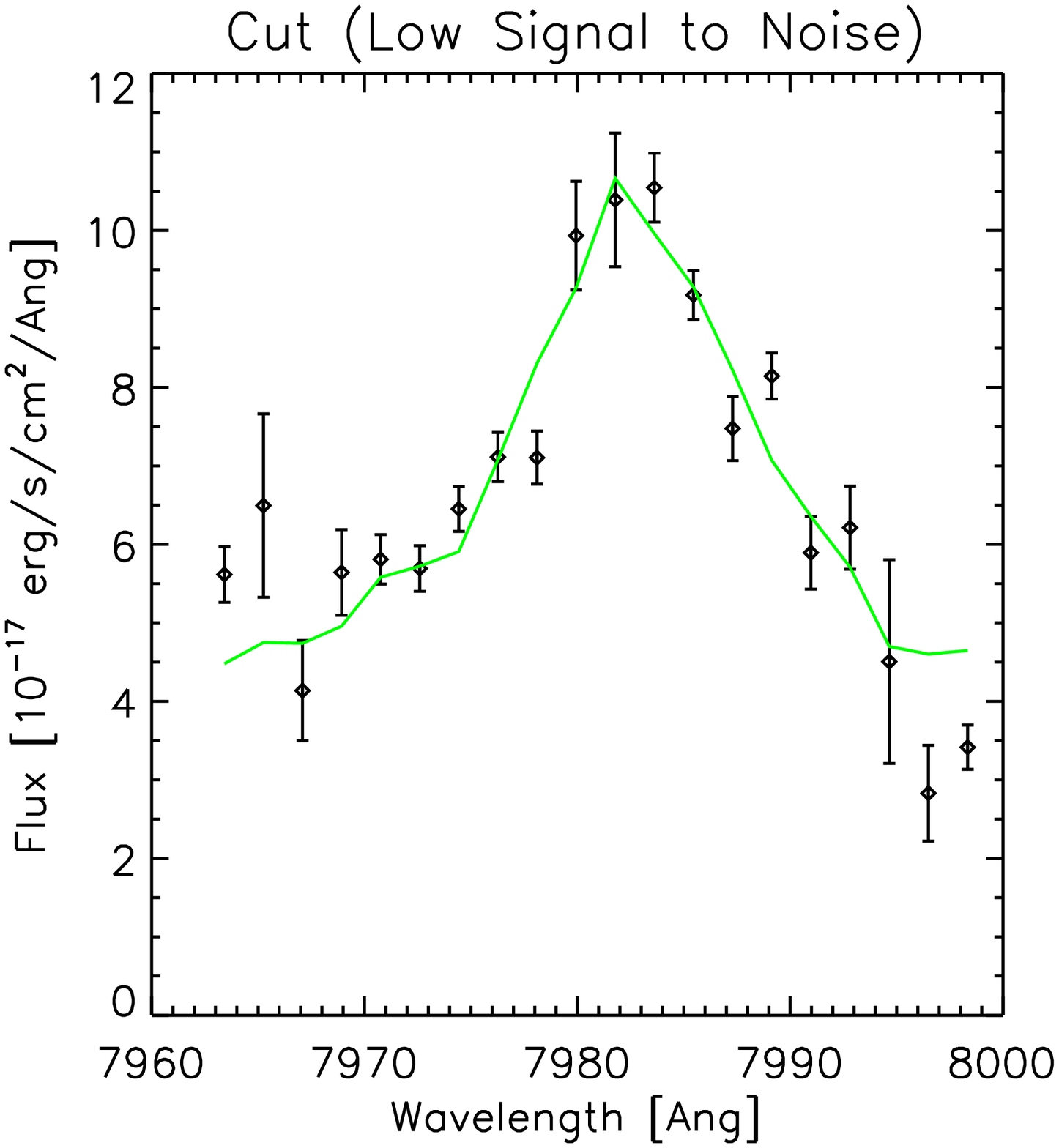,width=3in}}
\caption[]{\baselineskip=12pt Examples of four failed fits. The quasar emission lines shown in this
figure are the same as are shown in Figure~\ref{fig:failed}. For each of the four examples of failed
spectra shown in Figure~\ref{fig:failed}, Figure~\ref{fig:failedfit} shows the corresponding best (but
failed) spline fit of the measured shape of the 5007~\AA\ line (continuous curve)to  the measured shape
(points with error bars) of the 4959~\AA\ line. \label{fig:failedfit}}
\end{figure}

Figure~\ref{fig:failed} and Figure~\ref{fig:failedfit} illustrate the reasons why we need to impose
selection criteria, or tests, to determine which quasar spectra can yield precise measurements of the
wavelengths of the redshifted [O~{\sc iii}] lines.  Figure~\ref{fig:failed} shows the measured spectrum shape
for four different quasars and Figure~\ref{fig:failedfit}shows, for the same four quasars, the best
spline fits of the 5007~\AA\ line shapes to the measured shape of the 4959~\AA\ line.  The various
panels show spectra in which the measurements are compromised by different problems. In the upper left
panels of Figure~\ref{fig:failed} and Figure~\ref{fig:failedfit}, the
[O~{\sc iii}] lines have somewhat
different shapes and in the upper right panels the peaks of the two lines can not be lined up precisely
by a linear shift in wavelengths. In the lower left panels, the H$\beta$ line is so strong that it might
distort the profile of the 4959~\AA\ line. The spectra in the lower right panels have too low a signal
to noise ratio to allow a precise measurement.

In the following subsections, we shall define more quantitatively the criteria for passing these four
tests. We have formulated the tests as computer algorithms that can be applied quickly and easily to
numerical spectra.

We describe in \S~\ref{subsec:kstest} the Kolmogorov-Smirnov test that we have applied to the emission
line shapes in order to select acceptable quasar spectra. In \S~\ref{subsec:liningup}, we describe a test
which requires that the peaks of the two emission lines essentially lie on top of each other when the
line profiles are optimally shifted and rescaled. We present in \S~\ref{subsec:hbeta} a test that
eliminates quasar spectra with a large H$\beta$ contamination of the 4959~\AA\ line. Finally, we describe
in \S~\ref{subsec:signaltonoise} a requirement that the area under the 5007~\AA\ line be measured with a
high signal-to-noise ratio. For brevity, it is convenient to refer to these tests as the KS, line peak,
H$\beta$, and signal-to-noise criteria. In \S~\ref{subsec:feII}, we show {\it post facto} that Fe~{\sc
ii} lines do not compromise our measurements of $\alpha^2(t)$.

\subsection{Kolmogorov-Smirnov test}
\label{subsec:kstest}

We represent the shapes of the emission line profiles by the seven measured flux values that are centered
on the line peak. We then use a Kolmogorov-Smirnov (KS) test to determine whether the seven flux values
centered on the peak of the 4959~\AA\ line shape are, after multiplying by a constant factor $A$ [see
eq.~(\ref{eq:defnA}) and related discussion] consistent with being drawn from the same distribution as the
seven flux values centered on peak of the 5007~\AA\ line shape. The spectrum in the upper left hand
panel of Figure~\ref{fig:failed} was rejected because the two [O~{\sc iii}] lines have slightly different
shapes. For an unknown reason (which could be an instrumental or measuring error or simply a
fluctuation), the 4959~\AA\ line has a slight extra contribution on the short wavelength side of the
line profile.

If the two line shapes are the same before being affected by noise and measuring errors, as required by
the analysis described in \S~\ref{sec:matching}, then the two lines will generally pass a KS test.
We set our level for acceptance such that there are only a few false negatives, i.e., spectra
incorrectly rejected. In principle, there could be some false positives since the KS test does not
require that the flux values match in wavelength as well as in intensity. In practice, false positives of
this kind  essentially never occur.

 We require that the two sets of 7 flux values be drawn from the
same distribution to a confidence level (C.L.) corresponding to $2\sigma$ (95\% C.L.). From our initial
sample of 313 quasars in the redshift range that allowed the
observation of the [O~{\sc iii}] lines, 13 quasars
failed this test but no other test. Five of the 13 quasars that failed the test had values of $\alpha^2$
that differed by at least $2\sigma$ from the sample average of $\alpha^2$. In the full EDR sample, 28
quasars failed only this test.  Out of the 260 quasars that passed the preliminary spectral quality test
described in \S~\ref{subsec:introsample}, only 80 passed the KS test. Out of the total sample of 702
objects with redshifts that permitted the measurement of the O~{\sc iii} lines, only 99 passed the KS test.

The requirement that the shapes  of the 5007~\AA\ and 4959~\AA\ lines be similar according to the KS
test is our most stringent test.

We show in \S~\ref{sec:differentsamples} that the results are essentially unchanged if we use the
central 11 flux values instead of the central 7 values (see row four of Table~\ref{tab:average} and
Table~\ref{tab:linearfit}) or if we omit the KS test entirely (see row
eight of Table~\ref{tab:average} and
Table~\ref{tab:linearfit}).

\subsection{Lining up the peaks}
\label{subsec:liningup}

We require that the peak of the 4959~\AA\ line and the peak of the shifted 5007~\AA\ line lie in the
same pixel, for the best-fit value of $\eta$.  The demand that the two peaks be lined up after the 4959~\AA\ line is
optimally shifted strengthens the constraint that the two shapes be the same. Since this test requires
that the peaks line up but does not constrain the shift $\eta$, the test enforces the requirement that
the shapes of the two lines be the same but does not constrain the possible values of $\alpha$.  In
principle, this test may occasionally reject otherwise acceptable spectra if the peak of one of the lines
lies very near the boundary between two pixels. In the initial sample, only two quasars failed this test
but no other test. Both of the quasars which failed were $2\sigma$ outliers. In the full EDR sample, eight
quasars failed only this test. Out of the 260 quasars that passed the preliminary spectral quality test
described in \S~\ref{subsec:introsample}, 152 passed the test of the lining up of the peaks.  From
the total sample of 702 objects with appropriate redshifts for
measuring O~{\sc iii} lines, 247 passed the line
peak test.

If we omit the line peak test, the sample size increases by nine objects, but the overall results are
unchanged (see row nine of Table~\ref{tab:average} and Table~\ref{tab:linearfit}).

\subsection{H$\beta$ contamination}
\label{subsec:hbeta}

 The H$\beta$ line, which is centered at 4861~\AA, can contaminate
the 4959~\AA\ line if H$\beta$ is too strong. Moreover, the presence of a strong H$\beta$ line could in
principle distort the continuum fit. Therefore, we require that the area under the H$\beta$ line be no
more than twice the area under the 5007~\AA\ line. In practice, this requirement ensures that the
contamination of the 4959~\AA\ line from the H$\beta$ line is less than the contamination from the
5007~\AA\ line. In our initial sample, three quasars failed only this test. In the full EDR sample, four
quasars failed only this test. None of the quasars that failed this test were $2\sigma$ outliers. Out of
the 260 (702) quasars that passed the preliminary spectral quality test (for which the quasar redshift
permitted the measurement of the O~{\sc iii} lines), 231 (333) passed the H$\beta$ test.

The results of the analysis are essentially unchanged if the H$\beta$ test is omitted entirely (see row
ten of Table~\ref{tab:average} and Table~\ref{tab:linearfit}).

\subsection{Signal-to-Noise Test}
\label{subsec:signaltonoise}

Our final test is designed to eliminate spectra that are too noisy to make an accurate measurement of
$\alpha^2$.  We require that the area under the 5007~\AA\ line be measured to an accuracy of $\pm 5$ \%
as determined using the errors estimated from the Spectro 2-D code (Burles \& Schlegel
2003\nocite{burlesschlegel}). In other words, we require that the signal-to-noise ratio for the line
intensity is at least 20:1.  Six quasars failed only this test in our initial sample; one of the failed
quasars was a $2\sigma$ outlier. In the full EDR sample, 12 quasars failed only this test. Out of the 260
(702) quasars that passed the preliminary spectral quality test (for
which the O~{\sc iii} lines could have
been measured), 105 (404) passed the signal-to-noise test.

 The results are essentially unchanged if the signal-to-noise test is
omitted entirely (see row eleven of Table~\ref{tab:average} and Table~\ref{tab:linearfit}).

\subsection{Fe~{\sc ii}}
\label{subsec:feII}

There are two Fe~{\sc ii} lines, 4923.9 \AA \, and 5018.4 \AA \,, that are sometimes observed in the
spectra of quasars in the vicinity of the O~{\sc iii} lines and  which might conceivably influence the
measurement  of the shapes of the O~{\sc iii} lines. These Fe~{\sc ii} lines are often weak in quasar
spectra, perhaps especially in spectra that exhibit strong O~{\sc iii} emission (see, e.g., Fig.~1 of
Boroson \& Green 1992\nocite{boroson} or Vanden Berk et al. 2001).

In principle, the KS test of the similarity of the shapes of the O~{\sc iii} emission lines should
eliminate any spectra in which the Fe~{\sc ii} lines distort the line shapes and produce  apparent shifts
in the line centers. However, we decided to verify quantitatively this conclusion.

We tried to identify spectra in our standard sample (see \S~\ref{sec:standard} below) in which the
Fe~{\sc ii} lines were visible. This turned out to be difficult because of the weakness of the 4923.9 \AA
\, and 5018.4 \AA \, lines. The reader can verify directly the weakness of the Fe lines by examining the
spectra of the quasars in our standard sample (see \hbox{http://www.sns.ias.edu/$\sim$jnb} [See Quasar
Absorption and Emission Lines/Emission Lines on this site] and \S~\ref{subsubsec:full}).

We were forced to look at an expanded region of the spectrum to search for other potential Fe lines (see,
for example, Fig.~3 and Table~10 of Sigut \& Pradhan 2003\nocite{sigut} and also the discussion in Netzer \& Wills
1983\nocite{netzer}) that might be associated with the 4923.9 \AA \, and 5018.4 \AA \, lines.

We formed a sub-sample of the 42 spectra standard sample, which consisted of 7 objects that plausibly
might contain observable Fe~{\sc ii} lines. We compared the value of \hbox{$<\alpha^2(t)>/\alpha^2(0)$}
computed for the Fe-free 35 spectra with the value of $<\alpha^2(t)>/\alpha^2(0)$ determined for the
entire sample of 42 objects and with the value found for the 7 objects that might contain an observable
amount of Fe~{\sc ii}. Using the procedure described in \S~\ref{sec:standard}, we found
$<\alpha^2(t)>/\alpha^2(0) = 1.0000 \pm 0.0003$ for the Fe-free sample (35 quasars), which is essentially
identical with the result given in equation~(\ref{eq:averagestandard}) of \S~\ref{sec:standard} for the
entire standard sample of 42 quasars. The result for the 7 objects that may have observable Fe~{\sc ii}
lines, $<\alpha^2(t)>/\alpha^2(0) =  1.0021 \pm 0.0012$, has a much larger uncertainty, but is consistent
with the value found for the Fe-free sample (and for the total standard sample,
eq.~\ref{eq:averagestandard}).

We conclude that the Fe~{\sc ii} lines do not bias our measurement of $<\alpha^2(t)>$.

\section{Standard Sample}
\label{sec:standard}

\begin{figure}[!t]
\centerline{\psfig{figure=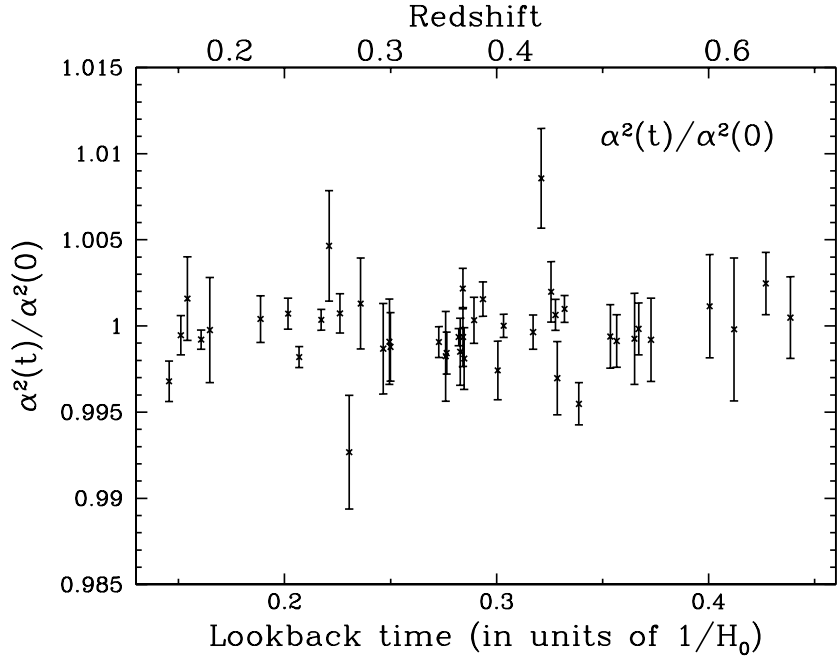,width=5in,angle=270}} \caption[]{\baselineskip=12pt The
fine-structure constant versus cosmic time. For our standard sample of 42 quasar spectra, the figure
shows the measured values of $\alpha(t)$ versus cosmic lookback time. The measurements are consistent
with a value of $\alpha$ that does not change with cosmic time. The lookback time is expressed in units
of $H_{0}^{-1} = 1.4\times 10^{10}\, {\rm yr^{-1}} $ [see eq.~(\ref{eq:defntime})].
\label{fig:standardalphavst}}
\end{figure}

In this section, we discuss the results of our analysis of the spectra of the 42 objects in our Standard
Sample. For the reader's convenience, we give in Table~\ref{tab:standard} of \S~\ref{appendix:B} of the
Appendix the values of the redshift, the measured quantity $\eta(z)$ [defined by
equation~(\ref{eq:defneta})] that determines $\alpha^2(z)$, and the relative scaling factor, $A$ (defined
in \S~\ref{sec:matching}), for all the QSO's in the Standard Sample.

Figure~\ref{fig:standardalphavst} shows, for the 42 standard QSO's, how the measured values of $\alpha$
depend upon cosmic lookback time. It is apparent from a chi-by-eye fit to
Figure~\ref{fig:standardalphavst} that there is no significant time dependence for the measured values of
$\alpha$ in our Standard Sample. The lookback time shown in Figure~\ref{fig:standardalphavst} is $H_0 t$,
which is independent of the numerical value of $H_0$ [see eq.~(\ref{eq:defntime})].

We present in \S~\ref{subsec:standardavplusslope} the average value of $\alpha$ measured for our
Standard Sample, as well as the best-fit slope, $\left<d\alpha(t)/dt\right>$. In
\S~\ref{subsec:bootstrap}, we describe how we have calculated the bootstrap errors for the sample
properties.
\subsection{Average and Slope for the Standard Sample}
\label{subsec:standardavplusslope}

The first row of Table~\ref{tab:average} shows that the weighted average value of $\alpha(z)$ for our
standard sample is

\begin{table}[!t]
\centering \caption[]{\baselineskip=12pt Average value of alpha. The table presents in column three the
measured weighted average value of $\alpha/\alpha(0)$  for our standard sample as well as the measured
average for 17 alternative cuts, defined in \S~\ref{subsec:alternatives}, on the data. The number of
quasar spectra that pass the cuts defining each sample is given in the second column. No alternative
sample produces an average value of $\alpha$ significantly different from the value obtained for the
standard sample.} \label{tab:average}
\begin{tabular}{ccc}
\tableline\tableline
Sample & Sample Size& Average $\alpha/\alpha(0)$\\
\tableline
Standard sample &                   42&     $ 1.00007 \pm 0.00014$\\
Unweighted errors &         42&     $ 1.00051 \pm 0.00022$\\
Strict H$\beta$ limit &         28&     $ 0.99993 \pm 0.00014$\\
Signal-to-noise of 10:1&        51&     $ 1.00009 \pm 0.00014$\\
11 point KS test &              68&     $ 1.00010 \pm 0.00013$\\
EW not area&                41&     $ 1.00005 \pm 0.00014$\\
$\chi^2$ instead of KS &        35&     $ 0.99991 \pm 0.00014$\\
Omit KS test &              70&     $ 1.00010 \pm 0.00013$\\
Omit peak line up &             52&     $ 1.00004 \pm 0.00013$\\
Omit H$\beta$ test &            45&     $ 1.00009 \pm 0.00014$\\
Omit signal-to-noise test &         54&     $ 1.00009 \pm 0.00014$\\
Remove worst outlier &          41&     $ 1.00006 \pm 0.00014$\\
$z<0.3632$ &                21&     $ 1.00005 \pm 0.00018$\\
$z>0.3632$ &                21&     $ 1.00011 \pm 0.00022$\\
$\Omega_m=1, \Omega_\Lambda=0$ &    42&     $ 1.00007 \pm 0.00014$\\
$\Omega_m=0, \Omega_\Lambda=1$ &    42&     $ 1.00007 \pm 0.00014$\\
Add $R(0)$ and $\sigma_{0}$ &       43&     $ 1.00002 \pm 0.00009$\\
Add $R(0)$ and $0.1\sigma_{0}$ &    43&     $ 1.00002 \pm 0.00009$\\
\tableline
\end{tabular}
\end{table}

\begin{equation}
\frac{\left<\alpha(z)\right>}{\alpha(0)} ~=~ 1.00007 \pm 0.00014 \, , \label{eq:averagestandard}
\end{equation}
where we have calculated each $\alpha^2(t)/\alpha^2(0)$ from equation~(\ref{eq:ratioofalphas}). We have used
the local value of $\alpha^2(0)$ [or, more precisely, $R(0)$, see eq.~(\ref{eq:Rofzero})] in
Table~\ref{tab:average} and in equation~(\ref{eq:averagestandard}) only to establish the scale.  We have not
made use of the local measurement of $R(0)$ in either Table~\ref{tab:average} or
equation~(\ref{eq:averagestandard}).

It is conventional to represent cosmological variations of $\alpha$ in terms of the average fractional
change over the time probed by the sample. In this language, the limit given in
equation~(\ref{eq:averagestandard}) is
\begin{equation}
\frac{\Delta \alpha}{\alpha(0)}~=~\left(-0.7 \pm 1.4\right) \times 10^{-4}.
\label{eq:deltaalphaoveralpha}
\end{equation}

 The average lookback time for our Standard Sample is
\begin{equation}
\left< t \right>~=~ \frac{0.28}{H_0} ~=~3.8\times 10^9 {\rm ~yr}\left(\frac{72~{\rm km
\,s^{-1}Mpc^{-1}}}{H_0}\right). \label{eq:lookback}
\end{equation}
Therefore, the characteristic variation of $\alpha$ permitted by our Standard Sample is
\begin{equation}
\frac{\Delta \alpha}{\left< t \right> \alpha} = \left( 2 \pm 4\right) \times 10^{-14} {\rm ~ yr^{-1}}.
\label{eq:characteristic}
\end{equation}
The mean (median) redshift for the Standard Sample is z = 0.37 (z= 0.37).

 The best-fit value of $d\alpha/\alpha \, dt$ can be obtained
by fitting the observed values of $\alpha^2(t)$ given in Table~\ref{tab:standard}  of
\S~\ref{appendix:B} of the Appendix to a linear relation, equation~(\ref{eq:defnlinear}), with a slope
defined by equation~(\ref{eq:defnslope}).  Using the slope given in Table~\ref{tab:linearfit}, we find for our
Standard Sample

\begin{equation}
\frac{1}{\alpha}\frac{d\alpha }{ dt} ~=~ \frac{H_0\times{\rm Slope}}{2}~=~\left(0.6 \pm 1.7\right) \times
10^{-13}\, {\rm yr^{-1}} \left(\frac{H_0}{72~{\rm km
\,s^{-1}Mpc^{-1}}}\right).\label{eq:alphadotstandard}
\end{equation}

The slope given in equation~(\ref{eq:alphadotstandard}) is obtained using only measurements made on the
spectra of the 42 quasars in our standard sample. No independent knowledge of $\alpha(0)$ is used. Hence,
the limit implied by equation~(\ref{eq:alphadotstandard}) is self-calibrating.

 In the first rows of Table~\ref{tab:average} and
Table~\ref{tab:linearfit}, we give the best-fit parameters, as well as their uncertainties, that were computed using the
bootstrap errors (see discussion of bootstrap errors in \S~\ref{subsec:bootstrap} below). In the
second rows of Table~\ref{tab:average} and Table~\ref{tab:linearfit}, we give the same quantities but
this time computed by assuming that all the errors are the same and equal to the average of the formal
measuring errors used for the first row calculations. The results are essentially the same.

\begin{table}[!t]
\centering \caption[]{\baselineskip=12pt Best linear fit. Here $\alpha^2(t) = \alpha^2_{\rm fit, 0}[1 +
H_0 S t]$, where the slope $S$ is defined by equation~(\ref{eq:defnslope}).
The value of the intercept, $\alpha^2_{\rm fit, 0} $
 and the slope  $S$ are both calculated from the quasar measurements summarized
in Table~\ref{tab:whowhat}. In the table, the value of $\alpha^2_{\rm fit, 0}/\alpha^2 ({\rm local ~
meas.})$ is defined by equation~(\ref{eq:ratioofalphas}). The time $t$ is calculated from
equation~(\ref{eq:defntime}) for a universe with the present composition of $\Omega_m=0.3$,
$\Omega_\Lambda=0.7$.} \label{tab:linearfit}
\begin{tabular}{cccc}
\tableline\tableline
Sample & Sample Size &$\alpha^2_{\rm fit, 0}/\alpha^2 ({\rm local ~ meas.})$& Slope $S$ \\
\tableline
Standard sample &                  42&     $ 1.0006 \pm 0.0012$&      $ -0.0016 \pm 0.0045$\\
Unweighted errors &        42&     $ 1.0005 \pm 0.0020$&      $  0.0020 \pm 0.0065$\\
Strict H$\beta$ limit&         28&     $ 0.9997 \pm 0.0013$&      $  0.0008 \pm 0.0050$\\
Signal-to-noise of 10:1  &     51&     $ 1.0004 \pm 0.0011$&      $ -0.0009 \pm 0.0043$\\
11 point KS test &         68&     $ 1.0012 \pm 0.0011$&      $ -0.0035 \pm 0.0038$\\
EW not area &              41&     $ 1.0006 \pm 0.0012$&      $ -0.0017 \pm 0.0045$\\
$\chi^2$ instead of KS &       35&     $ 1.0010 \pm 0.0011$&      $ -0.0043 \pm 0.0044$\\
Omit KS test &             70&     $ 1.0012 \pm 0.0010$&      $ -0.0034 \pm 0.0037$\\
Omit peak line up &        52&     $ 0.9999 \pm 0.0011$&      $  0.0005 \pm 0.0041$\\
Omit H$\beta$ test &           45&     $ 1.0006 \pm 0.0011$&      $ -0.0017 \pm 0.0044$\\
Omit signal-to-noise test &    54&     $ 1.0003 \pm 0.0011$&      $ -0.0004 \pm 0.0043$\\
Remove worst outlier &         41&     $ 1.0006 \pm 0.0012$&      $ -0.0016 \pm 0.0045$\\
$z<0.3632$ &               21&     $ 1.0009 \pm 0.0022$&      $ -0.0029 \pm 0.0097$\\
$z>0.3632$ &               21&     $ 1.0021 \pm 0.0047$&      $ -0.0056 \pm 0.0149$\\
$\Omega_m=1, \Omega_\Lambda=0$ &   42&     $ 1.0005 \pm 0.0011$&      $ -0.0013 \pm 0.0037$\\
$\Omega_m=0, \Omega_\Lambda=1$ &   42&     $ 1.0004 \pm 0.0007$&      $ -0.0005 \pm 0.0015$\\
Add $R(0)$ and $\sigma_{0}$ &      43&     $ 1.0002 \pm 0.0008$&      $ -0.0003 \pm 0.0030$\\
Add $R(0)$ and $0.1\sigma_{0}$ &   43&     $ 1.0002 \pm 0.0008$&      $ -0.0003 \pm 0.0030$\\
\tableline
\end{tabular}
\end{table}

\subsection{Bootstrap estimate of uncertainties}
\label{subsec:bootstrap}

The errors quoted in this section and elsewhere in the paper represent $1\sigma$ uncertainties calculated
using a bootstrap technique with replacement. For readers interested in the details, this is exactly what
we did to find the uncertainties in the quantities calculated for the standard sample. We created $10^5$
simulated samples by drawing 42 objects at random and with replacement from the real sample. For each of
these simulated realizations,  we calculate a weighted average of the sample. We then determine the
average and the standard deviation from the distribution, which was very well fit by a Gaussian, of the
weighted averages of the $10^5$ simulated samples.

We demonstrate in \S~\ref{sec:differentsamples} (cf.~rows 1 and 2 of Table~\ref{tab:average}  and
Table~\ref{tab:linearfit}) that the sample averages for the time-dependence of $\alpha$ are rather
insensitive to the estimated sizes of the individual error bars. By construction, the inferred
uncertainties only depend on the relative sizes of the assigned error bars in the observed sample.  The
estimated errors in the standard sample could all be multiplied by an arbitrary scale factor without
affecting the final inferred uncertainty for the sample average. Even the relative errors are not very
important in the present case, because all the errors are rather similar in the standard sample.

\section{Results for 18 Different Samples}
\label{sec:differentsamples}

Table~\ref{tab:average} and Table~\ref{tab:linearfit} present the results obtained for 17 samples of
quasar spectra in addition to our Standard Sample. These additional samples were defined by modifying in
various ways the selection criteria, described in \S~\ref{sec:tests}, that were used to select the
Standard Sample. The last one of the 17 alternate samples shown in Table~\ref{tab:average} and
Table~\ref{tab:linearfit} contains a hypothetical data point; this sample includes a local measurement of
$\alpha^2$ that is ten times more accurate than the best currently-available local measurement.

The 16 additional data samples in rows 2--17 of Table~\ref{tab:average} and Table~\ref{tab:linearfit} were
created from the SDSS data sample using less restrictive selection criteria. The sample presented in row
18 was created in order to determine whether a much improved measurement of $R(0)$ would significantly
affect the precision with which the time-dependence of $\alpha$ can be
investigated using O~{\sc iii}
measurements.

These additional data samples were created in order to test the robustness of our conclusions regarding
the lack of time dependence of $\alpha$, conclusions that are stated in \S~\ref{sec:standard}, especially
equation~(\ref{eq:deltaalphaoveralpha}) and equation~(\ref{eq:alphadotstandard}).

We answer here the following question: Does the lack of measured time dependence for $\alpha$ depend upon
how we define the data sample? The answer is no. Our conclusions regarding the time dependence of
$\alpha$ are robust; the conclusions are the same for all the variations we have made on the selection
criteria that define the samples.

All of the data samples that appear in equation~(\ref{eq:defnslope}) include only quasar spectra that pass
plausible tests, i.e., all of the included results are obtained with high-quality spectra. The reader
may be interested to know what the results would be if we calculated $\alpha^2(t)$ for the entire sample
of 260 quasars for which the O~{\sc iii} lines are measurable (see discussion at the end of
\S~\ref{subsubsec:spectra}), regardless of how badly individual spectra fail the quality tests.

If we indiscriminately use all 260 quasar spectra, we find a sample average
\begin{equation}
<\alpha^2(t)>/\alpha^2(0) = 1.00040 \pm 0.00026~~({\rm indiscriminate~sample}) \, .
\label{eq:indiscriminately}
\end{equation}
If we force the 260 measured values of $\alpha^2(t)$ to fit a linear function of cosmic time, we find
$\alpha^2_{\rm fit}(0) = 1.0003 \pm 0.0011$ and slope $S = 0.0003 \pm  0.0040$. The slope $S$ is defined
by equation~(\ref{eq:defnslope}).  We calculate the uncertainties for this indiscriminate sample using
unweighted bootstrap errors (cf. \S~\ref{subsec:bootstrap}). By comparison with the values obtained for
samples that pass the quality tests, given in Table~\ref{tab:average} and Table~\ref{tab:linearfit}, we
see that the errors are about four times larger for the indiscriminate sample although the results are in
satisfactory agreement with the values obtained using only spectra that pass plausible tests.

We begin in \S~\ref{subsec:alternatives} by describing the 17 samples, all of which pass some set of
quality tests, that we have studied in addition to the Standard Sample. We discuss in
\S~\ref{subsec:numericalalternative} the numerical results obtained for the alternative samples.
\subsection{Definitions of Alternative Samples}
\label{subsec:alternatives}

In this section, we describe briefly how the 17 non-standard data samples were determined. In the
comments below, we only state how the defining criteria are changed for the alternative samples. For each
alternative sample, each  test discussed in \S~\ref{sec:tests} that is not mentioned explicitly
below has been used in selecting the alternative sample. The descriptions are ordered in the same way as
the samples appear in Table~\ref{tab:average} and Table~\ref{tab:linearfit}.

\begin{description}
\item[Unweighted errors.] This sample is created from the data of the Standard
Sample by artificially setting all of the errors to be equal.
\item[Strict H$\beta$ limit.] The H$\beta$ limit used in defining
this sample is twice as strict as for the Standard Sample. The area under the H$\beta$ emission line is
required to be less than or equal to the area under the [O~{\sc iii}] 5007~\AA\ line. This stronger
requirement removes 14 of the 42 quasars in the standard sample.
\item[Signal-to-noise of 10:1.] For our standard sample we require that the area under the 5007~\AA\ emission be
measured to an accuracy of 5\%, corresponding to a signal-to-noise ratio of 20:1. We relax this
signal-to-noise criterion to 10:1 for this alternative, which increases the standard sample by an
additional 9 quasars.
\item[11 point KS test.] Eleven points centered on the peaks of the
[O~{\sc iii}] 4959~\AA\ line and the 4007~\AA\ line are used in carrying out the KS test, instead of the 7
points used in defining the standard sample. This change in the number of points used increases the
sample size by more than 50\%, from 42 to 68 quasars.
\item[Equivalent widths rather than areas.] It is common in
astronomy to describe emission lines in terms of equivalent width, the number of Angstroms of continuum
flux that is equal to the total flux in an emission line. We have required for this alternative sample
that the uncertainty in the measured equivalent width
 of the 5007~\AA\ line be no more than 5\%, i.e.,
that the signal to noise ratio of the measured equivalent width be 20:1. For the Standard Sample, the
20:1 requirement is made on the area under the emission line, not on the equivalent width. Replacing the
requirement on the accuracy of the measurement of the area under the 5007~\AA\  emission line by a
requirement on the equivalent width removes one of the 42 quasars from
our standard sample. We note that the average
(median) equivalent width of the 5007~\AA\ line in the Standard Sample is 68~\AA\ (57~\AA).
\item[$\chi^2$ instead of KS.] For this sample, we replaced the KS test with a $\chi^2$ test. We used the
20 pixel region of the spectrum centered on the expected position of the 4959~\AA\ line and compared the
shape of this line with the best-fit shifted and rescaled shape of the 5007~\AA\ line. We require that
$\chi^2$ per degree of freedom be less than one.  We obtain a total sample size of 35 if we use the
$\chi^2$ test instead of the KS test.
\item[Omit KS test.] We now describe the results of eliminating, one at a time, each one of our criteria
for inclusion in the standard sample. Our most stringent test is the KS test. If we omit this test
entirely, the alternative sample is increased by 28 objects to a total
sample
of 70 quasars. This is the
largest sample we consider.
\item[Omit peak line up.] The requirement that the peaks of the 5007~\AA\ and 4959~\AA\ lines line up
within one pixel may occasionally eliminate some good spectra. We have omitted the lining-up requirement,
which increases the standard sample by an additional 10 quasars.
\item[Omit H$\beta$ test.] For the standard sample, we have placed a stringent requirement on the possible
H$\beta$ contamination. If we omit the H$\beta$ test entirely, the sample is increased by 3 quasars.
\item[Omit signal-to-noise test.] We omit entirely the signal-to-noise test, increasing the sample by 12
additional objects.
\item[Remove worst outlier.] We remove from the standard sample the quasar that has a value of $\alpha^2$
that differs the most from the average value of $\alpha^2$ within the standard sample. This object is
$3.8\sigma$ from the standard sample average.
\item[Require $z < 0.3632$.] We split the standard sample into two equal parts in order to test whether
$\alpha^2$ is the same in the low and the the high redshift halves of the sample. This sample is the low
Z half of the standard sample.
\item[Require $z > 0.3632$.] This is the high redshift half of the standard sample.
\item[$\Omega_m = 1.0$, $\Omega_{\Lambda} = 0.0$.] To test for the sensitivity of our results to our
assumptions regarding cosmological parameters, we recalculated all quantities using cosmological times
appropriate for an extreme universe with $\Omega_m = 1.0$, $\Omega_{\Lambda} = 0.0$.
\item[$\Omega_m = 0.0$, $\Omega_{\Lambda} = 1.0$.] We also analyzed the data using the
extreme cosmological parameters $\Omega_m = 0.0$, $\Omega_{\Lambda} = 1.0$.
\item[Add $R(0)$ and $\sigma(0)$.] We added to the standard sample the local measurement summarized in
equation~(\ref{eq:locallambdas})-equation~(\ref{eq:Rofzero}).
\item[Add $R(0)$ and $0.1\sigma(0)$.] This sample is hypothetical. We supposed that the
accuracy of the local measurement could be improved by a factor of 10. The purpose of studying this
sample was to investigate whether a better local measurement would improve the over-all accuracy of the
determination of the time dependence of $\alpha$.
\end{description}

We have not considered samples in which a precise external value of $\alpha$ is introduced, a value of
$\alpha$ determined by measurements that do not involve the O~{\sc iii} emission lines. In order to use
external values of $\alpha$, we would need to rely upon an extremely accurate theoretical calculation of
the fine structure splitting for the two O~{\sc iii} lines. This reliance would depart from  the empirical
approach we have adopted; it would subject our analysis to some of the same concerns that we
express in \S~\ref{sec:comparison} with respect to the Many-Multiplet method. We limit ourselves
to considering ratios of O~{\sc iii} wavelengths [see eq.~(\ref{eq:defnR}) and eq.~(\ref{eq:ratioofalphas})],
since many recognized systematic uncertainties (and perhaps also some unrecognized uncertainties) are
avoided in this way.
\subsection{Numerical results for Alternate samples}
\label{subsec:numericalalternative}

We discuss in this subsection the numerical results for the 17 alternative samples including their
average values (\S~\ref{subsubsec:averages}), their slopes
(\S~\ref{subsubsec:slopes}), the influence of
$R(0)$ (\S~\ref{subsubsec:local}), and the insensitivity of the sample errors to the estimates of
the errors for individual measurements of $\alpha^2$ (\S~\ref{subsubsec:errors}).

\subsubsection{Average values}
\label{subsubsec:averages}

 Table~\ref{tab:average} shows that all of the samples we have considered yield average values for
 $\alpha^2$ that lie within the range
 \begin{equation}
 \frac{\left<\alpha\right>}{\alpha(0)} ~=~ 1.00007^{+0.00044}_{-0.00016} \, ,
\label{eq:average18}
 \end{equation}
where the fiducial value of ${\left<\alpha^2\right>}/{\alpha^2(0)} = 1.00007$ was chosen because it is
the average value for the standard sample.  Since the calculated errors on the average of each sample are
 between $\pm 0.0001$ and $\pm 0.0002$, it is clear from equation~(\ref{eq:average18}) that all of the samples
 shown in Table~\ref{tab:average} give consistent values for $\left<\alpha\right>/\alpha(0)$.

 The results given in Table~\ref{tab:average} are all consistent with a fine-structure constant that does
 not vary with time. The characteristic variation permitted is
 $\Delta \alpha/{\left< t \right> \alpha} = \left( 2 \pm 4 \right) \times 10^{-14} {\rm ~ yr^{-1}}$ [see
 eq.~(\ref{eq:characteristic})].

 \subsubsection{Slopes}
 \label{subsubsec:slopes}

Table~\ref{tab:linearfit} gives the best-fitting slopes, S, for an assumed linear dependence of
$\alpha^2(t)$ on $t$. All of the slopes are consistent with zero within $2\sigma$. There is no
significant evidence for a time dependence of $\alpha$ in any of the samples we have considered. Using
the quoted value of $H_0$ to convert the slopes, S, given in Table~\ref{tab:linearfit} to time
derivatives (and including the factor of two that relates the slopes of $\alpha^2(t)$ and $\alpha(t)$),
we conclude that [cf. eq.~(\ref{eq:alphadotstandard})]

\begin{equation}
\left|\frac{1}{\alpha}\frac{d\alpha}{dt}\right| ~<~ 2\times 10^{-13} {\rm \, yr^{-1}} ~.
\label{eq:slopes}
\end{equation}

\subsubsection{The role of the local value of $R(0)$}
\label{subsubsec:local}

The various samples for which results are given in the first 16 rows of Table~\ref{tab:average} and
Table~\ref{tab:linearfit} do not include any local measurements of $R$ [cf. eq.~(\ref{eq:defnR})].  The
last two rows of Table~\ref{tab:average} and Table~\ref{tab:linearfit} contain an additional data point
representing the local value of $R(0)$, as presented in equation~(\ref{eq:Rofzero}).

We see from rows 17 and 18 of Table~\ref{tab:average} and Table~\ref{tab:linearfit} that adding the local
value of $R(0)$ does not change significantly either the sample average of $R(0)$ or the slope describing
the time dependence. Even if one assumes, as was done in constructing row 18 of Table~\ref{tab:average}
and Table~\ref{tab:linearfit}, that the uncertainty in the local value of $R(0)$ could be reduced by a
factor of ten, there would still not be a significant impact  on the accuracy with which the time
dependence of $\alpha^2$ would be known.  The existing error on $R(0)$
is already much less than the other errors that enter our analysis.

\subsubsection{Errors}
\label{subsubsec:errors}

The bootstrap method that we have used to calculate the errors is insensitive to the average size of the
errors made in individual measurements of $\alpha^2$ and depends only weakly on the relative errors (see
the discussion of bootstrap errors in
 \S~\ref{sec:standard}).  The robustness of the bootstrap
errors can be seen most clearly by comparing the results for rows one and two in Table~\ref{tab:average}
and Table~\ref{tab:linearfit}.  In row one, the bootstrap uncertainty was estimated for the standard
sample making use, in computing an average for each of the $10^5$ simulated samples, of the actual errors
determined from our individual measurements. In row two, the bootstrap uncertainty was estimated by
assuming that the errors on all measurements were identical. The results shown in rows one and two of
Table~\ref{tab:average} and Table~\ref{tab:linearfit} are essentially identical for both the sample
average and the time dependence of $\alpha$.

\section{Comparison of O~{\sc iii} method with Many-Multiplet method}
\label{sec:comparison}

In this section, we compare the O~{\sc iii} emission line method for studying the time-dependence of the
fine-structure constant with what has been called the `Many-Multiplet' method.   The Many-Multiplet
method is an extension of, or a variant on, previous absorption line studies of the time-dependence of
$\alpha$. We single out the Many-Multiplet method for special discussion since among all the studies done
so far on the time dependence of the fine-structure constant, only the results obtained with the
Many-Multiplet method yield statistically significant evidence for a time dependence. All of the other
studies, including precision terrestrial laboratory measurements (see references in Uzan
2003)\nocite{uzan} and previous investigations using quasar absorption lines (cf. Bahcall, Sargent, \&
Schmidt 1967; Wolfe, Brown, \& Roberts 1976; Levshakov 1994; Potekhin \& Varshalovich 1994, Cowie \&
Songaila 1995; and Ivanchik, Potekhin, \& Varshalovich 1999) or AGN emission lines (Savedoff 1956;
Bahcall \& Schmidt 1967), are consistent with a value of $\alpha$ that is independent of cosmic time. The
upper limits that have been obtained in the most precise of these previous absorption line studies are
generally $|\Delta \alpha/\alpha(0)|~< ~ 2 \times 10^{-4}$, although Murphy et al.
(2001c)\nocite{murphyad} have given a limit that is ten times more restrictive. None of the previous
absorption line studies has the sensitivity that has been claimed for the Many-Multiplet method.

We first describe in \S~\ref{subsec:MMmethod} the salient features of the Many-Multiplet method.
Then in \S~\ref{subsec:contrasts} we compare and contrast the Many-Multiplet method for measuring
the time-dependence of $\alpha$ with the O~{\sc iii} method.

\subsection{The Many-Multiplet method}
\label{subsec:MMmethod}

In the last several years, a number of papers have discussed the ``Many-Multiplet" method for determining
the time dependence of $\alpha$ using quasar absorption line spectra (see Dzuba, Flambaum, \& Webb
1999a,b;\nocite{dzuba99a,dzuba99b} Webb et al. 1999;\nocite{webb99} Murphy et al.
2001a,b;\nocite{murphy01a,murphy01b} and Webb et al. 2002, as well as references quoted therein). Because
the Many-Multiplet method uses many atomic transitions from different multiplets in different ionization
stages and in different atomic species, some of the transitions having large relativistic corrections,
the Many-Multiplet method has a potential sensitivity that is greater than the sensitivity that has been
achieved with the O~{\sc iii} method or any other astronomical method. The Many-Multiplet collaboration
has inferred a change in alpha with cosmic epoch, $\Delta \alpha/\alpha(0)~=~(-0.7 \pm 0.2)\times
10^{-5}$, over a redshift range from z = 0.2 to 3.5 (see especially Murphy et al. 2001a,b and Webb et al.
2002 and references to earlier in this paragraph).

 The Many-Multiplet method compares the measured wavelengths of absorption (not emission)
lines seen in quasar spectra with the measured or computed laboratory wavelengths of the same lines.
Lines from the ions Mg~{\sc i}, Mg~{\sc ii}, Al~{\sc ii}, Si~{\sc ii},
Cr~{\sc ii}, Fe~{\sc ii}, Ni~{\sc ii},and Zn~{\sc ii} are used (cf. Table~1 of Murphy et
al. 2001b). In addition, different multiplets from the same ion are used.  The lower redshift (smaller
lookback time) results are based upon measurements made using Mg~{\sc
i}, Mg~{\sc ii}, and Fe~{\sc ii}, while the larger
redshift results use lines observed from Al~{\sc ii}, Al~{\sc iii},
Si~{\sc ii}, Cr~{\sc ii}, Ni~{\sc ii}, and Zn~{\sc ii}.

In order to increase the measuring precision, the Many-Multiplet (see
Murphy et al. 2001b) ``...is based
on  measuring the wavelength separation between the resonance transitions of {\it different ionic
species} with no restriction on the multiplet to which the transitions belong."

As emphasized by Murphy et al.~2001b, the absolute laboratory values of the wavelengths of many
individual transitions must be precisely known from terrestrial experiments and measured precisely in the
quasar spectra in order to reach a higher precision than is possible by measuring the ratios of
fine-structure splittings to average wavelengths. In the O~{\sc iii} method, and in the many applications
of doublet splittings of absorption lines that originate on a single ground state, the relativistic terms
appear only in the relatively small fine-structure splittings of the excited states of the atoms. The
Many-Multiplet method utilizes the fact that atomic ground states have larger relativistic, i.e.,
$\alpha^4$, contributions to their energies than do excited states. Instead of concentrating on
splittings, in which the ground state contributions cancel out, the Many-Multiplet method requires and
makes use of the measurement of individual wavelengths.

Sophisticated theoretical atomic physics calculations are required in order to determine how changing the
value of $\alpha$ affects the wavelengths of the different transitions employed in the Many-Multiplet
method. In particular, one must calculate the derivatives of the transition wavelengths with respect to
$\alpha^2$.  Therefore, Dzuba, Flambaum, \& Webb (1999a,b) have developed new theoretical tools to
calculate the relativistic shift for each transition of interest. They use a relativistic Hartree-Fock
method to construct a basis set of one-electron orbitals and a configuration interaction method to obtain
the many-electron wave function of valence electrons. Correlations between core and valence electrons are
included by means of many-body perturbation theory.

The known systematic uncertainties have been discussed extensively in a series of papers by the
originators of the Many-Multiplet method. These comprehensive and impressive studies cover many different
topics. It is far beyond the scope of this article to discuss in detail the systematic uncertainties and
the accuracy obtainable with the Many-Multiplet method. The reader is referred to an excellent
presentation of the Many-Multiplet method in a series of published papers (see especially Murphy et al.
2001a; as well as Dzuba, Flambaum, \& Webb 1999a,b; Webb et al. 1999;\nocite{webb99} Murphy et al. 2001b;
 Webb et al. 2002; and the recent preprint by Webb, Murphy, Flambaum, \& Curran 2003\nocite{webb02}).

\subsection{Contrasts between the O~{\sc iii} method and the Many-Multiplet method}
\label{subsec:contrasts}

In this subsection, we  compare briefly some practical consequences of  two strategies for measuring the
time dependence of $\alpha$, the O~{\sc iii} method and the Many-Multiplet method.  In the papers by the
Many-Multiplet collaboration, there are extensive discussions of the advantages of that method so we will
not repeat their arguments here. The reader is referred to the original papers of the Many-Multiplet
group for an eloquent description of the virtues of their method
(Dzuba, Flambaum, \&
Webb 1999a,b; Webb
et al. 1999; Murphy et al. 2001a,b; and Webb et al. 2002).

Table~\ref{tab:comparison} shows a schematic outline of the differences between the two methods.
\begin{table}[!t]
\centering \caption[]{\baselineskip=12pt Comparison of the O~{\sc iii} and Many-Multiplet
Methods.\label{tab:comparison}}
\begin{tabular}{lll}
\tableline\tableline
&O~III&Many-Multiplet\\
\tableline
Number of ions&one&many; different $Z,A$\\
Transitions&one&many\\
Multiplets&one&many\\
Velocity structure&identical&assumed: all ions same\\
Misidentification&no&a concern\\
Theory&simple&relativistic, many body\\
Full disclosure& yes & extremely difficult\\
\tableline
\end{tabular}
\end{table}
\subsubsection{Ions, transitions, multiplets}
\label{subsubsec:ions}

 For the O~{\sc iii} method, only one ion with one pair of transitions from the same
multiplet is used.

For the Many-Multiplet method, one uses many lines from many multiplets arising from ions with different
atomic numbers and abundances (and in some cases, difference ionization stages).  At the large redshifts
for which the Many-Multiplet collaboration has reported a slightly smaller value of the fine-structure
constant, the Many-Multiplet sample contains about 2 times as many lines in about 0.5 as many quasar
spectra as we include in our standard sample (cf.~Table 1 of Murphy et al.~2001a with
Table~\ref{tab:standard} of this paper).  The quoted error of the Many-Multiplet collaboration is
typically two orders of magnitude smaller than our quoted error.

\subsubsection{Velocity structure}
\label{subsubsec:lineprofiles}

\subsubsubsection{Velocity structure for the O~{\sc iii} method} \label{subsubsubsec:oiiilineprofile}

 The emission line profile is identical for all transitions used in
the O~{\sc iii} method. Indeed, this is one of the four principal criteria employed to select the O~{\sc
iii} sample (see \S~\ref{subsec:kstest}).

\subsubsubsection{Velocity structure for the Many-Multiplet method} \label{subsubsubsec:mmmlineprofile}

 Murphy et al. (2001b) have stressed that for the
Many-Multiplet method ``The central assumption in the analysis is that the velocity structure seen in one
ion corresponds exactly to that seen in any other ion. That is, we assume that there is negligible proper
motion between corresponding velocity components of all ionic species."   No such assumption is required
for the O~{\sc iii} method.

Thus the Many-Multiplet technique requires that, for each absorption line redshift, the likelihood
function be maximized with respect to a set of velocity structures (velocity profiles) that represent the
different individual cloud velocities that contribute to each of the absorption lines. The number of
velocity components and their relative velocities are not known a priori and must be solved for in the
maximization process. Typically, five or more independent cloud velocities are required to fit the data.
In their first paper suggesting a time-dependent fine-structure constant, the Many-Multiplet
collaboration noted conservatively that an anomaly in their data might be explained by "... additional
undiscovered velocity components in the absorbing gas" Webb et al. 1999).

How  nearly identical do the line profiles have to be in order not to produce an apparent variation of
$\alpha$ when none is present?

 How large would the difference between velocity profiles (relative component intensities)
have to be in order to mimic with a constant value of $\alpha$ the effect observed by the Many-Multiplet
collaboration? We can make an approximate estimate of the required velocity difference between different
ions by considering a simplified model in which only two lines from two different ions are observed, e.
g., one line from Cr~{\sc ii} and one line from Fe~{\sc ii}. Using a formalism similar to that described
by Murphy et al.~2001b, we write the frequency of line $i$ that is observed as an absorption line
originating at redshift z as

\begin{equation}
\omega_z(i) = (1 + z)\left[\omega_0(i) + q(i)x \right]\, , \label{eq:defnomegaz}
\end{equation}
where $x \equiv \alpha^2(z)/\alpha^2(0) - 1 \sim 2(\Delta \alpha/\alpha)$ and $q(i) = (1 +
z)^{-1}\partial \omega_z(i)/\partial x$. In terms of the quantities $q_1$ and $q_2$ given in Table~1 of
Murphy et al. 2001b, $q \approx q_1 + 2q_2$. The equations for absorption lines 1 and 2 that correspond
to equation~(\ref{eq:defnomegaz}) can be solved simultaneously for x, which measures the difference
between $\alpha(0)$ and $\alpha(z)$.  One finds

\begin{equation}
x ~=~ \left[\frac{\omega_z(1)\omega_0(2) - \omega_z(2)\omega_0(1)}{\omega_z(2)q(1) -
\omega_z(1)q(2)}\right]\, . \label{eq: simultaneous}
\end{equation}
If  $\alpha$ is a constant and the sources of the two absorption lines have identical velocity structures
as assumed in the Many-Multiplet method,  then $\omega_z(1) = (1 + z) \omega_0(1)$ and $\omega_z(2) = (1
+ z) \omega_0(2)$, and equation (\ref{eq: simultaneous}) yields $x \equiv 0$.

Now suppose that we keep $\Delta \alpha/\alpha= 0$ but allow one of the ions to have a non-zero velocity
$v$ with respect to the other ions, i. e.,

\begin{equation}
\omega_{z,~{\rm observed}}(1) = \omega_{z,~{\rm true}}(1)\left(1 + v/c\right) \, . \label{eq:vdefn}
\end{equation}
Then if one solves equation~(\ref{eq: simultaneous}) for the best-fit value of $x$, or $\Delta
\alpha/\alpha$, one finds an apparent value for $\Delta \alpha/\alpha$ of

\begin{equation}
\left(\frac{\Delta \alpha}{\alpha}\right)_{\rm apparent} ~=~\left(\frac{v}{2c}\right)\frac{\omega_{\rm
z,~true}(1)\left[\omega_0(2) + q(2)x\right]}{\left[\omega_z(2)q(1) - \omega_{\rm z,~true}(1)
q(2)\right]}. \label{eq:alphaapparent}
\end{equation}

Thus if one ignores the presence of the relative velocity, $v$, between the different ions that produce
the two absorption lines and solves for the best-fit value of $x$, one finds, as shown in
equation~(\ref{eq:alphaapparent}), an apparently non-zero value of $\Delta \alpha/\alpha$ even though by
hypothesis $\alpha$ was set equal to a constant.

We can invert equation~(\ref{eq:alphaapparent}) and solve for the magnitude of the relative velocity $v$
that is required to produce an apparent variation in $\Delta \alpha/\alpha = -7 \times^{-6}$ (see Murphy
et al.~2001b) 2 even if one is not present.  We consider as examples the strongest absorption lines of
Fe~{\sc ii} ($ \lambda = 2382.7$ \AA\,, $q = 1638 {\rm ~cm^{-1}}$), Cr~{\sc ii} ($\lambda = 2056.3$
\AA\,, $q = -1056 {\rm ~cm^{-1}}$), and Mg~{\sc ii} ($\lambda = 2796.4$ \AA\,, $q = 211 {\rm ~cm^{-1}}$),
where we have taken the values of $\lambda$ and $q$ from Table~1 of  Murphy et al.~2001b. Since all of
these lines have comparable wavelengths, they could all appear in the same sample of observed absorption
lines.

For all three pairwise combinations of the Fe~{\sc ii}, Cr~{\sc ii}, and Mg~{\sc ii} lines,  we find,
using equation~(\ref{eq:alphaapparent}), that the relative velocity $v$ is related to the apparent change
in $\alpha$ by the equation
\begin{equation}
v_{\rm relative} ~\approx~ 0.1 \left(\Delta \alpha/\alpha\right)_{\rm apparent}c.
\label{eq:vtodeltaalpha}
\end{equation}
Thus a relative velocity between a pair of the absorbing ions as small as $\sim 0.2~{\rm km \, s^{-1}}$
could give rise to the apparent change in $\alpha$ claimed by the Many-Multiplet collaboration. The
characteristic range of absorption velocities included within a single absorption system is three orders
of magnitude larger, i. e., of order $2\times 10^3 ~{\rm km \, s^{-1}}$.

 Webb, Murphy, Flambaum, and Curran 2003\nocite{webb02} have argued that any effect due to the
difference i2n velocity profiles (velocity structures) of ions from different elements will average out
if a large number of QSO absorption systems (or different sight-lines) are included in the sample. Are
there enough sight-lines available for this averaging to work to the required accuracy?

There are four ions with a large sensitivity to $\alpha^2$ that are listed in Table~1 of Murphy et
al.~2001b; these ions are Zn~{\sc ii} (2 potentially identifiable absorption lines), Cr~{\sc ii} (3
potential lines), Fe~{\sc ii} (7 potential lines), and Ni~{\sc ii} (3 potential lines). Of the 21
high-redshift absorption line systems listed in Table~2 of Murphy et al.~2001b, all five of the Zn~{\sc
ii} and Cr~{\sc ii} lines are identified in 4 separate absorption complexes. In three of these systems,
all 8 of the potentially observable strong  Ni~{\sc ii}, Zn~{\sc ii}, and Cr~{\sc ii}  lines are
identified.

It seems reasonable to wonder if these several very plausible absorption line systems constitute a
sufficiently large number of very reliably identified systems, systems in which all of the potentially
observable strong lines are detected (cf. discussion of line identifications in
\S~\ref{subsubsec:linestrengths}), for random effects to average out to an accuracy of $0.2~{\rm km \,
s^{-1}}$ over a velocity range of more than $10^2~{\rm km \, s^{-1}}$.

The Many-Multiplet assumption that all ions have the same velocity profile may be testable directly by
different groups using very high resolution absorption-line spectroscopy (resolution $10^5$ or better) on
relatively bright quasars. If the assumption of essentially identical line profiles is correct, one would
expect that all absorption lines measured by the Many-Multiplet collaboration at a given redshift would
be proportionally represented in all sub-clouds. Since absorption structures break up at high resolution
into many clouds (see Bahcall 1975\nocite{bahcall75} or almost any modern high-resolution spectroscopic
study of quasar absorption line spectra) , one can compare the relative strengths of the Mg~{\sc i},
Mg~{\sc ii}, Al~{\sc ii}, Si~{\sc ii}, Cr~{\sc ii}, Fe~{\sc ii}, Ni~{\sc ii}, and Zn~{\sc ii} lines in
different sub-clouds. One might possibly be able to use the different measured line strengths to
construct composite line profiles for different ions.

\subsubsection{Misidentification}
\label{subsubsec:linestrengths}

\subsubsubsection{In the O~{\sc iii} method} \label{subsubsubsec:oiiimisid}

For the O~{\sc iii}  method, we consider only strong O~{\sc iii} emission lines, lines that can easily be
recognized by a computer (or by a human eye).  We include only  O~{\sc iii} lines that have a high
signal-to-noise ratio, at least 20:1 (see \S~\ref{subsec:signaltonoise}). In our Standard Sample, the
average (median) equivalent width of the 5007~\AA\ line is 68~\AA\ (57~\AA). The interested reader can
inspect the reduced spectra for all of the quasars in our Standard Sample; the spectra are available at
http://www.sns.ias.edu/~jnb (See Quasar Absorption and Emission Lines/Emission Lines on this site).

There is no significant chance that the O~{\sc iii} lines will be misidentified or strongly blended.
There is no other pair of strong emission lines in the relevant part of quasar (or galaxy) spectra
(Vanden Berk et al. 2001\nocite{vandenberk}). We algorithmically  exclude spectra in which the H$\beta$
line is sufficiently strong to possibly contaminate the measurement of the wavelengths of the 5007~\AA\
and 4959~\AA\  lines.

\subsubsubsection{In the Many-Multiplet method} \label{subsubsubsec:manymultipletmisid}

For the Many-Multiplet method, line identifications   could be a source of previously unrecognized
systematic uncertainties .

In the only Many-Multiplet sample for which some of the identifications can be checked (see Table~4 of
Murphy et al. 2001b), there are three  examples in which the weaker line of the Al~{\sc iii} doublet
(1854.7\AA\, f = 0.7; 1862.8, f = 0.54 \AA\ ) is reported to be present but the stronger line of the
Al~{\sc iii} doublet is not observed (see in Table~4  the entries for QSO 0201+37 at $z_{abs} = 1.956$
and $z_{abs} = 2.325$ and QSO 1759+75 at $z_{\rm abs} = 2.625$). There are also three  examples in which
the weakest line of the Cr~{\sc ii} triplet (2056.3, f = 0.11; 2062.2, f = 0.08; and 2066.2, f = 0.05) is
reported as identified while the two stronger lines of the Cr~{\sc ii} triplet are not observed (see in
Table~4 the entries for QSO 0201+37 at $z_{\rm abs} = 1.956$ and $z_{\rm abs} = 2.462$ and QSO 1759+75 at
$z_{\rm abs} = 2.625$).

These seemingly unphysical identifications raise questions regarding the validity of the identification
procedures used by the Many-Multiplet collaboration. The identification software with which we are
familiar automatically rejects cases in which the weaker (or weakest) line in a multiplet is identified
in the absence of the strong lines (see, e. g., Bahcall 1968 or Bahcall et al. 1993, 1996). In all six of
the cases cited in the previous paragraph, we can verify from Table~4 of Murphy et al. 2001b that the
missing strong lines are in the observed region of the spectrum since observed lines are reported at
wavelengths above and below the missing lines.

Line identification codes for quasar absorption lines are necessarily complicated and the logic of how
different lines are identified depends, among other things, the decision tree used when multiple
identifications are possible (e.g., at different redshifts) for the same absorption lines, the confidence
level at which one makes the line identifications,  the assumed standard lines, the physical assumptions
that one makes about the absorbers (e.g., chemical composition and ionization state),  as well as  the
signal-to-noise ratio of each spectrum, and  the richness of the absorption line spectrum (for a
discussion of some of these complications see, e.g., the discussion of the first absorption-line code in
Bahcall~1968\nocite{bahcall68} and also Bahcall et al.~1993, 1996\nocite{bahcall93,bahcall96}).

The Many-Multiplet collaboration has not described in any detail their logic for their reported
identifications, the decision tree that was used, the confidence level that was adopted, the assumed
standard lines and the chemical composition, or the physical assumptions. The detailed spectra have also
not been published. It is therefore not possible to assess how large a contribution the misidentification
of lines could be to the systematic uncertainty in the evaluation of $\alpha(z)$ by the Many-Multiplet
collaboration. However, the fact noted above that the identification procedure used by the Many-Multiplet
collaboration results in weak lines being identified while stronger lines in the same multiplet are not
observed suggests that the systematic uncertainty due to misidentifications might be significant.

\subsubsection{Theory}
\label{subsubsec:theory}

 For the O~{\sc iii} method, only general theoretical considerations are required to determine the
dependence of $\alpha$ upon the measured quantity. All that is needed is the recognition that the fine
structure splitting is smaller than the non-relativistic atomic energy by a quantity that is proportional
to $\alpha^2$. As explained in \S~\ref{sec:outline}, the time dependence of $\alpha^2(t)$ can be
determined by measuring the ratio of the difference of two wavelengths divided by their sum [see
eq.~(\ref{eq:ratioofalphas})].

The situation is different for the Many-Multiplet method. Sophisticated many-body relativistic
calculations (Dzuba, Flambaum, \& Webb 1999a,b), including electronic correlations, are required in order
to determine the way the measured wavelengths depend upon $\alpha$ for the many different lines used in
the Many-Multiplet method.

\subsubsection{Wavelength Measurements}
\label{subsubsec:wavelengths}

 The O~{\sc iii} method only makes use of the difference and the ratio of
wavelengths, while the Many-Multiplet method requires knowing accurately the  individual values of the
wavelengths for all the transitions that are used. Some systematic uncertainties that are important if
one uses the individual values of the wavelengths (needed for the Many-Multiplet method) cancel out when
one considers only ratios of wavelengths (as appropriate for the O~{\sc iii} method).

For the O~{\sc iii} method, we have achieved a sample average accuracy for the wavelength splitting of
$0.01$~\AA, corresponding to an accuracy of individual measurements of about $0.07$~\AA. This accuracy
corresponds to individual measurement errors of slightly better than a tenth of a pixel, which is a
relatively modest precision for the high signal-to-noise spectra we have used.

The Many-Multiplet method requires more precise wavelength measurements. However, a  precise estimate of
the required accuracy can only be made by the Many-Multiplet collaboration since the collaboration has
not published the detailed data required to make an accurate calculation. These data include the number
of lines and the average relativistic splitting that were measured in each absorption system shown in, e.
g., Figure~3 of Murphy et al. 2001b.

\subsubsection{Full disclosure}
\label{subsubsec:full}

All of the spectra we use in measuring the O~{\sc iii} lines are available in the SDSS EDR~(see Schneider
et al. 2002, see also Richards et al. 2002)\nocite{richards02}. We describe in detail in
\S~\ref{sec:outline}---\S~\ref{sec:tests} how our measurements were made. Table~\ref{tab:whowhat} lists
all of the quasars that passed at least three of the four standard tests we have used.

Reduced spectra of the region around the O~{\sc iii} emission lines are available publicly for all
quasars that are included in our standard sample. The O~{\sc iii} lines stand out clearly  (to visual
inspection or to algorithmic selection). The spectra can be down-loaded at
\hbox{http://www.sns.ias.edu/$\sim$jnb}.

The situation is more complex for the Many-Multiplet method. Not all of the spectra are publicly
available. The line identifications depend upon many assumptions (see discussion in
\S~\ref{subsubsec:linestrengths} and in  Bahcall~1968\nocite{bahcall68} and Bahcall et al.~1993,
1996\nocite{bahcall93,bahcall96}). One must decide whether a given set of lines is, e.g., Zn~{\sc ii} at
one redshift or a combination of lines from different ions at different redshifts. Using spectra taken
over a large redshift range ($0.2 < z <3.7$) with different telescopes, full disclosure under these
circumstances would be an enormous task. Consistent with this difficulty, the detailed reduced spectra
have not made been made public by the Many-Multiplet collaboration and, for example, the justifications
for the line identifications of less abundant ions like Cr~{\sc ii}, Ni~{\sc ii}, and Zn~{\sc ii} are not
given explicitly.

If full disclosure for the Many-Multiplet method is not feasible, what would be necessary for an outside
researcher to form some impression of the validity of the line identifications? One would want to see a
published list of all of the absorption lines measured in each quasar spectrum with a quantitative
assessment of the strength and reliability of each line. Then one would like to know what was the
complete set of standard (unredshifted) absorption lines used to make identifications. Next, one would
like to see for all the absorption lines,  not just those with high sensitivity to $\alpha^2$, the
suggested absorption line identifications. One would like to answer elementary questions like: are the
theoretically strongest lines identified with the observed strongest lines?  Are there alternative
identifications at different redshifts?  Finally, one would like to see the results of Monte Carlo
simulations that address the issues of non-uniqueness of line identifications. Given the importance of
the subject, it seems to us that this degree of disclosure, which was exceeded for example in the Key
Project for Quasar Absorption Lines (e.g., Bahcall et al. 1993, 1996, and Januzzi et al.
1998)\nocite{jannuzi98}, is justified although it would require considerable additional effort.

\section{Discussion}
\label{sec:discussion}

We use the  strong nebular lines of O~{\sc iii}, 5007~\AA\ and 4959~\AA,  to establish an upper limit to
the time dependence of the fine structure constant, $|\alpha^{-1}(d\alpha/dt)| ~<~ 2 \times 10^{-13} {\rm
\, yr^{-1}}$ [see eq.~(\ref{eq:slopes})]. The limit is essentially the same for all 17 variations of the
selection criteria for the sample that we have analyzed. Even if we use all of the 260 spectra in which
the O~{\sc iii} lines are measurable, omitting all quality cuts on the data, we obtain  results for
$<\alpha(t)>/\alpha(0)$ and $|\alpha^{-1}(d\alpha/dt)|$ that are consistent with the results obtained for
our standard sample with four standard quality cuts on the data. Of course, for the 'indiscriminate'
sample with a total of 260 spectra, the estimated uncertainty is larger than for the smaller standard
(purified) sample [compare eq.~(\ref{eq:indiscriminately}) with eq.~(\ref{eq:averagestandard})].

The lack of a measurable time dependence can be seen visually in Figure~\ref{fig:standardalphavst}.  We
have used spectra from the early data release sample (EDR, Schneider et al. 2002) of the Sloan Digital
Sky Survey. {\bf Note added in proof.} We have analyzed the much larger sample of quasars recently made
available in the SDSS Data Release One sample (DR1, Schneider et al. 2003)\nocite{DR1} using the same
technique and code described in the main body of the present paper. We present our results for the SDSS
Data Release One sample in \hbox{Appendix C} The results obtained from the SDSS EDR and the SDSS DR1
samples are in good agreement with each other, but the inferences from the DR1 sample have smaller errors
(consistent with which is expected statistically from the larger number of quasars included in the DR1
sample) .

The reduced spectra for the 42 quasars in our standard sample are available publicly at the URL
\hbox{http://www.sns.ias.edu/$\sim$jnb}. The O~{\sc iii} lines stand out strongly in all spectra that we
measure.

The upper limit derived here on the change of $\alpha(t)$ over a characteristic time of $4\times 10^9$ yr
[cf. eq.~(\ref{eq:lookback})] is robust. We present in \S~\ref{sec:differentsamples} and in
Table~\ref{tab:average} and Table~\ref{tab:linearfit} results for 17 different algorithms for determining
the average value, and the rate of change, of $\alpha$ during the cosmic epoch explored. The results for
all the samples we have considered are in agreement.

Depending upon the algorithm adopted for selecting the sample, the sample size varies from a minimum of
28 quasars to a maximum of 70 quasars, with 42 quasars in our Standard Sample (cf.
Table~\ref{tab:average} and Table~\ref{tab:linearfit}). Essentially identical results are obtained if one
uses a weighted or unweighted average of the measurements, adopts a more stringent restriction on the
strength of possibly contaminating H$\beta$ emissions, relaxes the signal to noise requirement, omits or
changes the way the Kolmogorov-Smirnov test is applied, omits each of the other three defining selection
criteria, considers equivalent widths rather than area
 under the O~{\sc iii} emission lines, removes the most
distant outlier, compares measurements made for small and large redshifts, adopts an extreme cosmology,
or includes the result for a local measurement of $\alpha$.

The principal results for the O~{\sc iii} method are independent of the precise value of the fine
structure constant at the present epoch.  No assumption was made about the precise value of $\alpha^2(0)$
in deriving the numerical constraints on the time dependence of $\alpha$ that are presented in the first
16 rows (samples) of Table~\ref{tab:average} and Table~\ref{tab:linearfit}. The results given in the
first 16 rows of the tables and in equation~(\ref{eq:alphadotstandard}) depend only on measurements of
the ratio of $\alpha^2(t)/\alpha^2(0)$. If one is only interested in whether or not
$\alpha^2(t)/\alpha^2(0)$ is time dependent, it does not matter what constant value one assumes for
$\alpha^2(0)$.  In this sense, the O~{\sc iii} method is self-calibrating.

As a by-product of the measurements performed here, we derived in \S~\ref{subsec:amplitudes} a precise
value, $2.99 \pm 0.02$, for the ratio of the photon fluxes corresponding to  the 5007~\AA\ and the
4959~\AA\ lines. Thus the ratio of the Einstein A coefficients for the two lines is
${A\left(5007\right)}/{A\left(4959\right)} = 2.99 \pm 0.02$. Our measured value for
${A\left(5007\right)}/{A\left(4959\right)}$ is in good agreement with previous theoretical estimates,
which provides some support for the validity of the measurements that we have made. However, our
measurement of ${A\left(5007\right)}/{A\left(4959\right)}$ is more accurate than the previous theoretical
estimates.

Over the immense cosmic time interval explored using the [O~{\sc iii}] emission lines, the fractional change in
$\alpha$ is small, $\Delta \alpha/\alpha(0) = (0.7 \pm 1.4) \times 10^{-4}$. This null result is
consistent with all the measurements that we know about from other methods (Savedoff 1956; Bahcall \&
Schmidt 1967; Bahcall, Sargent, \& Schmidt 1967; Wolfe, Brown, \& Roberts 1976; Levshakov 1994; Potekhin
\& Varshalovich 1994,  Cowie \& Songaila 1995; Ivanchik, Potekhin, \& Varshalovich 1999; and Uzan 2003),
including other measurements of quasar absorption line spectra and terrestrial laboratory measurements.
Only the recent results that were obtained using the Many-Multiplet method suggest a non-zero
time-dependence for $\alpha$; the suggested rate of change is consistent with the previous and present
null results for astronomical measurements. The Many-multiplet
collaboration concludes in published papers that $\Delta
\alpha/\alpha(0)~=~(-0.7 \pm 0.4)\times 10^{-5}$ over the redshift range of 0.5 to 3.5 (Murphy et al.
2001b; see also Dzuba, Flambaum, \& Webb 1999a,b; Webb et al. 1999; Murphy et al. 2001a; and Webb et al.
2002).

Table~\ref{tab:comparison} and \S~\ref{sec:comparison} compare the O~{\sc iii} and the Many-Multiplet
method.  The O~{\sc iii} method is simpler and less subject to systematic uncertainties, but the
Many-Multiplet method has the advantage of greater potential precision since one considers many more
transitions. As explained in \S~\ref{sec:comparison}, the higher precision claimed for the Many-Multiplet
method comes at a price, namely: 1) the necessity for assuming that the line profiles of different ions
are identical even though the lines are formed in a number of different clouds; and 2) the possible
misidentification of absorption features with absorption redshifts that are not known {a priori}.
However, nothing that we have said in this paper contradicts the claims by the Many-Multiplet
collaboration to have measured a time-dependence at a level of sensitivity well below what is currently
obtainable with the O~{\sc iii} method. We have not found, nor even looked for, an error in the analysis
of the Many-Multiplet collaboration, although we have listed some concerns in \S~\ref{sec:comparison}.
The assumption by the Many-Multiplet collaboration that the velocity profiles (velocity structures) of
different ions are the same may be testable with independent, very high-resolution spectra of relatively
bright quasars (see discussion in \S~\ref{subsubsec:lineprofiles}).

The Many-Multiplet collaboration finds a small, non-zero but approximately constant value for $\Delta
\alpha/\alpha(0)$ above redshifts of about unity. The effect reported by the Many-Multiplet collaboration
is one and a half orders of magnitude smaller than the upper limit obtained in this paper using the Early
Release SDSS data. We believe that, for the O~{\sc iii} method, statistical errors dominate the
uncertainty in the current measurements. We would like to improve the result given in this paper by using
much larger quasar samples from the SDSS and 2dF surveys, when they become publicly available.  The
accuracy of the O~{\sc iii} results can also be improved by using higher-resolution spectra obtained with
large telescopes.

\acknowledgments We are grateful to Dr. J. Reader, Group Leader of Atomic Spectroscopy at the National
Institute of Standards and Technology, for valuable suggestions and advice. The reader will find useful
related information at the NIST Atomic Spectra Database:
http://physics.nist.gov/cgi-bin/AtData/main$_{-}$asd. We are indebted to B. Draine, G. Holder, D. Maoz,
M. T. Murphy, J. Schaye, and M. Strauss for valuable comments on drafts of this paper (including the
first draft that was posted on astro-ph) and to R. Lupton and M. Strauss for generous help with data
reduction issues. We appreciate helpful email exchanges with V. V. Flambaum, M. T. Murphy, and J. Webb.
JNB is supported in part by a NSF grant \#PHY-0070928.

\clearpage

\appendix
\section{Is R(t) proportional to $\alpha^2(t)$?}
\label{appendix:A}

The principal assumption made in the text is that the difference in wavelengths divided by their sum is
proportional to the fine-structure constant squared. For simplicity of description, we have proceeded as
if the numerator of $R(t)$ [see eq.~(\ref{eq:roftagain})] were proportional to $\alpha^4$ and the
denominator was proportional to $\alpha^2$, so that their ratio was proportional to $\alpha^2$. In fact,
we know that there are both $\alpha^2$ and $\alpha^4$ terms (non-relativistic and relativistic terms) in
the denominator.

How do the presence of both $\alpha^4$ and $\alpha^2$ terms in the denominator of $R(t)$ [cf.
eq.~(\ref{eq:defnR})] affect the slope, S [cf. eq.~(\ref{eq:defnslope})], which represents the linear time
dependence of $\alpha^2(t)$? We answer this question in this Appendix.

Let the shorter wavelength, higher frequency transition be denoted by $\nu_1$. Then we can write
\begin{equation}
\nu_1 ~=~ \nu_0\left[1 + B_1\alpha^2\right],\,\,\nu_1 ~=~
\frac{c}{4959~{\rm \AA}}\,, \label{eq:defnnu1}
\end{equation}
where $\nu_0$ ($\propto \alpha^2$) is the non-relativistic atomic energy difference and the relativistic
terms, including the spin orbit interaction, are represented by $B_1 \alpha^2$. Similarly, for the longer
wavelength, lower frequency transition we can write
\begin{equation}
\nu_2 ~=~ \nu_0\left[1 + B_2 \alpha^2\right],\,\,\nu_2 ~=~
\frac{c}{5007~{\rm \AA}}. \label{eq:defnnu2}
\end{equation}

The measured fine-structure splitting, $(\Delta \nu)_{\rm fine~structure}$, satisfies the relation
\begin{equation}
(\Delta \nu)_{\rm fine~structure} ~\propto~ \left(B_1 - B_2\right)\alpha^2(t).\label{eq:finestructure}
\end{equation}
In principle, we could use relativistic Hartree-Fock calculations to estimate the values of $B_1$ and
$B_2$.  We prefer not to use theoretical calculations that could have an unknown systematic error and
instead rely upon measured wavelengths only. This empiricism results in a small  ambiguity in the meaning
of any time dependence that is ultimately measured. However, as we shall now show, this ambiguity is
numerically unimportant.

The quantity we use to measure $\alpha^2(t)/\alpha^2(0)$ is [cf. eq.~(\ref{eq:defnR})]
\begin{equation}
\frac{R(t)}{R(0)} ~=~ \left[\frac{\lambda_2(t) - \lambda_1(t)}{\lambda_2(t) + \lambda_1
(t)}\right]\left[\frac{\lambda_2 (0) + \lambda_1(0)}{\lambda_2(0) - \lambda_1 (0)}\right]\,.
\label{eq:roftagain}
\end{equation}
Substituting equation~(\ref{eq:defnnu1}) and equation~(\ref{eq:defnnu2}) into equation~(\ref{eq:roftagain}) and carrying
out the straightforward algebra, we find
\begin{equation}
\frac{R(t)}{R(0)}  ~\cong~ \left[1 + S tH_0\left\{1 - \frac{\left(B_1 +
B_2\right)\alpha^2(0)}{2}\right\}\right] \,,\label{eq:rcorrection}
\end{equation}
where the slope $S$ is defined by equation~(\ref{eq:defnslope}). Let $\Delta \nu$ be the energy difference
corresponding to the wavelength splitting of $47.93~\AA\ $ (cf. Fig.~\ref{fig:oiii}). Then one can
easily see that if $\nu_2 < \nu_0 < \nu_1$, then
\begin{equation}
\left|\frac{\left(B_1 + B_2\right)\alpha^2(0)}{2}\right| ~\sim~ \frac{\Delta\nu}{2\nu_0} \simeq 0.005 \,.
\label{eq:approx}
\end{equation}
If $\nu_0$ does not satisfy the above inequality, then $|\left(B_1 + B_2\right)\alpha^2(0)/2|$ can be of
order $0.01$.

In all cases, the presence of both $\alpha^2$ and $\alpha^4$ terms in the denominator of $R(t)$ only
changes the interpretation of the slope $S$ by a negligible amount. The next higher order terms in the
atomic energies (or frequencies), which arise from the Lamb shift,  change the interpretation of $S$ by
less than 1\%.

\section{Data Tables}

In this section, we present data tables that may be of interest to the specialist.
Table~\ref{tab:standard} presents the measurements of $\eta$ and $A$, and the inferred value of
$\alpha^2/\alpha^2(0)$,  for the Standard Sample of 42 quasars. Table~\ref{tab:whowhat} presents, for all
95 quasars that passed at least three of the four standard selection tests, the tests that each quasar
passed and the measured values of $\eta$ and $A$.  Table~\ref{tab:alternatives} lists the SDSS name and
an alternative catalog name for quasars we have studied that have been previously listed in other
catalogs.

 \label{appendix:B}
\begin{deluxetable}{lcccc}
\tablewidth{0pt} \tablecaption{\baselineskip=12pt Measurements for the Standard Sample.  The table
contains the quasar name (with an asterisk if previously discovered by another survey), the SDSS
redshift, the measured value of $\alpha^2$, the relative displaced $\eta$ of the $5007$~\AA\  and
$4959$~\AA\  lines, and the relative scaling $A$, of the two lines. Quasars marked with an asterisk have
an alternative catalog name listed in Table~\ref{tab:alternatives}. \label{tab:standard}} \tablehead{
\colhead{Quasar Name}&\colhead{$z$}&\colhead{$\alpha^2/\alpha^2(0)$}&\colhead{$\eta \times
10^3$}&\colhead{$A$}} \startdata
SDSS 000859+011351 & 0.28678 & 0.4824$\pm$0.0013 & 9.6953 & 2.78 \\
SDSS 001030+010006 & 0.37792 & 0.4819$\pm$0.0005 & 9.6853 & 3.04 \\
SDSS 001327+005231* & 0.36219 & 0.4844$\pm$0.0011 & 9.7345 & 2.96 \\
SDSS 001545+000822 & 0.37091 & 0.4814$\pm$0.0006 & 9.6738 & 3.15 \\
SDSS 005717+003242 & 0.49178 & 0.4811$\pm$0.0009 & 9.6694 & 2.99 \\
SDSS 011420$-$004049 & 0.43913 & 0.4803$\pm$0.0011 & 9.6526 & 3.18 \\
SDSS 012050$-$001832 & 0.34901 & 0.4810$\pm$0.0012 & 9.6656 & 2.97 \\
SDSS 012750$-$000919 & 0.43766 & 0.4813$\pm$0.0004 & 9.6734 & 2.96 \\
SDSS 013352+011345* & 0.30804 & 0.4807$\pm$0.0010 & 9.6610 & 3.04 \\
SDSS 015950+002340* & 0.16313 & 0.4824$\pm$0.0010 & 9.6944 & 3.28 \\
SDSS 020115+003135 & 0.36229 & 0.4807$\pm$0.0008 & 9.6613 & 3.06 \\
SDSS 021259$-$003029 & 0.39449 & 0.4811$\pm$0.0004 & 9.6676 & 2.94 \\
SDSS 021359+004226 & 0.18243 & 0.4809$\pm$0.0003 & 9.6637 & 2.93 \\
SDSS 021652$-$002335 & 0.30452 & 0.4846$\pm$0.0014 & 9.7395 & 2.82 \\
SDSS 024706+002318 & 0.36322 & 0.4799$\pm$0.0006 & 9.6447 & 2.97 \\
SDSS 025432$-$004220* & 0.43390 & 0.4822$\pm$0.0007 & 9.6907 & 3.03 \\
SDSS 032559+000800 & 0.36019 & 0.4826$\pm$0.0013 & 9.6996 & 2.84 \\
SDSS 033202$-$003739 & 0.60729 & 0.4834$\pm$0.0021 & 9.7158 & 3.06 \\
SDSS 034106+004610 & 0.63359 & 0.4802$\pm$0.0007 & 9.6508 & 2.70 \\
SDSS 095625$-$000353* & 0.51217 & 0.4815$\pm$0.0008 & 9.6763 & 2.93 \\
SDSS 102700$-$010424 & 0.34379 & 0.4809$\pm$0.0005 & 9.6652 & 3.00 \\
SDSS 104132$-$005057 & 0.30281 & 0.4823$\pm$0.0014 & 9.6918 & 3.24 \\
SDSS 105151$-$005117* & 0.35892 & 0.4805$\pm$0.0002 & 9.6571 & 2.99 \\
SDSS 111353$-$000217 & 0.44544 & 0.4813$\pm$0.0004 & 9.6723 & 2.99 \\
SDSS 115758$-$002220 & 0.25960 & 0.4811$\pm$0.0004 & 9.6684 & 2.99 \\
SDSS 130002$-$010601 & 0.30732 & 0.4809$\pm$0.0004 & 9.6648 & 3.02 \\
SDSS 130713$-$003601 & 0.17012 & 0.4808$\pm$0.0006 & 9.6627 & 2.96 \\
SDSS 134113$-$005315* & 0.23745 & 0.4836$\pm$0.0007 & 9.7190 & 2.91 \\
SDSS 134459$-$001559* & 0.24475 & 0.4806$\pm$0.0005 & 9.6579 & 3.01 \\
SDSS 134507$-$001900 & 0.41867 & 0.4808$\pm$0.0005 & 9.6618 & 3.05 \\
SDSS 135553+001137 & 0.45959 & 0.4822$\pm$0.0021 & 9.6902 & 3.05 \\
SDSS 142648+005323 & 0.21956 & 0.4805$\pm$0.0007 & 9.6572 & 2.96 \\
SDSS 145221+002359 & 0.45792 & 0.4790$\pm$0.0005 & 9.6263 & 3.13 \\
SDSS 150629+003543 & 0.36980 & 0.4831$\pm$0.0012 & 9.7086 & 2.91 \\
SDSS 172032+551330 & 0.27235 & 0.4813$\pm$0.0005 & 9.6735 & 2.95 \\
SDSS 172206+565451 & 0.42554 & 0.4847$\pm$0.0013 & 9.7407 & 2.78 \\
SDSS 173311+535457 & 0.30721 & 0.4811$\pm$0.0010 & 9.6678 & 3.25 \\
SDSS 173602+554040 & 0.49675 & 0.4811$\pm$0.0018 & 9.6677 & 2.83 \\
SDSS 173638+535432 & 0.40808 & 0.4814$\pm$0.0006 & 9.6750 & 3.20 \\
SDSS 234145$-$004640 & 0.52386 & 0.4822$\pm$0.0011 & 9.6914 & 3.04 \\
SDSS 235439+005751 & 0.38974 & 0.4799$\pm$0.0009 & 9.6451 & 2.84 \\
SDSS 235441$-$000448 & 0.27877 & 0.4777$\pm$0.0013 & 9.6000 & 2.72 \\
\enddata
\end{deluxetable}

\begin{deluxetable}{lcccc}
\tablewidth{0pt} \tablecaption{\baselineskip=12pt Results for 95 quasars. Table~\ref{tab:whowhat}
includes all SDSS EDR quasars that passed at least three of the four standard tests described in
\S~\ref{sec:tests}. The table shows which of the four tests the quasar passed. The tests (a)-(d),
refer, respectively to the KS test, lining up the peaks of the two
[O~{\sc iii}] emission lines, the H$\beta$
contamination, and the signal to noise ratio of the lines. The quantity $\eta$ is defined by
equation~(\ref{eq:defneta}) and $A$ is the ratio of the intensity of the 5007~\AA\ line to the intensity of
the 4959~\AA\ line.  Quasars marked with an asterisk have an alternative catalog name listed in
Table~\ref{tab:alternatives}.  \label{tab:whowhat}} \tablehead{
\colhead{Quasars}&\colhead{$z$}&\colhead{Tests}&\colhead{$\eta\times 10^{+3}$}&\colhead{A}} \startdata
SDSS 000859+011351 & 0.28678 & (a)(b)(c)(d) & 9.6953 & 2.78 \\
SDSS 001002$-$010107 & 0.55477 & (a)~~~~(c)(d) & 9.8746 & 3.73 \\
SDSS 001030+010006 & 0.37792 & (a)(b)(c)(d) & 9.6853 & 3.04 \\
SDSS 001327+005231* & 0.36219 & (a)(b)(c)(d) & 9.7345 & 2.96 \\
SDSS 001545+000822 & 0.37091 & (a)(b)(c)(d) & 9.6738 & 3.15 \\
SDSS 003431$-$001312 & 0.38095 & ~~~~(b)(c)(d) & 9.6948 & 3.09 \\
SDSS 004458+004319 & 0.34979 & ~~~~(b)(c)(d) & 9.6679 & 2.87 \\
SDSS 005717+003242 & 0.49178 & (a)(b)(c)(d) & 9.6694 & 2.99 \\
SDSS 005905+000651* & 0.71858 & ~~~~(b)(c)(d) & 9.6496 & 2.78 \\
SDSS 011420$-$004049 & 0.43913 & (a)(b)(c)(d) & 9.6526 & 3.18 \\
SDSS 012050$-$001832 & 0.34901 & (a)(b)(c)(d) & 9.6656 & 2.97 \\
SDSS 012750$-$000919 & 0.43766 & (a)(b)(c)(d) & 9.6734 & 2.96 \\
SDSS 013352+011345* & 0.30804 & (a)(b)(c)(d) & 9.6610 & 3.04 \\
SDSS 015629+000724 & 0.36025 & (a)(b)~~~~(d) & 9.6916 & 2.74 \\
SDSS 015950+002340* & 0.16313 & (a)(b)(c)(d) & 9.6944 & 3.28 \\
SDSS 020115+003135 & 0.36229 & (a)(b)(c)(d) & 9.6613 & 3.06 \\
SDSS 021123+001959 & 0.48699 & (a)(b)~~~~(d) & 9.6924 & 3.11 \\
SDSS 021225+010056 & 0.51265 & ~~~~(b)(c)(d) & 9.6420 & 3.03 \\
SDSS 021259$-$003029 & 0.39449 & (a)(b)(c)(d) & 9.6676 & 2.94 \\
SDSS 021359+004226 & 0.18243 & (a)(b)(c)(d) & 9.6637 & 2.93 \\
SDSS 021652$-$002335 & 0.30452 & (a)(b)(c)(d) & 9.7395 & 2.82 \\
SDSS 022119+005628 & 0.40007 & (a)~~~~(c)(d) & 9.6225 & 2.86 \\
SDSS 022259$-$005035 & 0.55255 & (a)(b)~~~~(d) & 9.6418 & 3.06 \\
SDSS 024240+005727* & 0.56896 & (a)~~~~(c)(d) & 9.6839 & 3.19 \\
SDSS 024706+002318 & 0.36322 & (a)(b)(c)(d) & 9.6447 & 2.97 \\
SDSS 024954+010148* & 0.58546 & (a)~~~~(c)(d) & 9.6876 & 2.86 \\
SDSS 025007+002525* & 0.19729 & (a)(b)(c)~~~~ & 9.6899 & 2.88 \\
SDSS 025432$-$004220* & 0.43390 & (a)(b)(c)(d) & 9.6907 & 3.03 \\
SDSS 025646+011349 & 0.17655 & ~~~~(b)(c)(d) & 9.6885 & 3.02 \\
SDSS 025735$-$001631 & 0.36242 & (a)(b)~~~~(d) & 9.6869 & 3.68 \\
SDSS 030048+005440* & 0.66171 & ~~~~(b)(c)(d) & 9.6171 & 2.74 \\
SDSS 030313$-$001457* & 0.70018 & ~~~~(b)(c)(d) & 9.7288 & 2.90 \\
SDSS 031226$-$003709 & 0.62124 & ~~~~(b)(c)(d) & 9.6550 & 2.95 \\
SDSS 032559+000800 & 0.36019 & (a)(b)(c)(d) & 9.6996 & 2.84 \\
SDSS 032628$-$002741 & 0.44535 & ~~~~(b)(c)(d) & 9.6812 & 2.43 \\
SDSS 033202$-$003739 & 0.60729 & (a)(b)(c)(d) & 9.7158 & 3.06 \\
SDSS 033606$-$000754 & 0.43163 & ~~~~(b)(c)(d) & 9.6669 & 3.12 \\
SDSS 034106+004610 & 0.63359 & (a)(b)(c)(d) & 9.6508 & 2.70 \\
SDSS 034247+010932 & 0.36003 & ~~~~(b)(c)(d) & 9.6978 & 3.00 \\
SDSS 034345$-$004801 & 0.74618 & (a)~~~~(c)(d) & 9.6385 & 3.17 \\
SDSS 094132+000731 & 0.48882 & (a)(b)~~~~(d) & 9.5457 & 3.09 \\
SDSS 094241+005652 & 0.69488 & ~~~~(b)(c)(d) & 9.6728 & 3.10 \\
SDSS 095625$-$000353* & 0.51217 & (a)(b)(c)(d) & 9.6763 & 2.93 \\
SDSS 101419$-$002834 & 0.35850 & (a)~~~~(c)(d) & 9.7623 & 3.43 \\
SDSS 102502+003126* & 0.36260 & ~~~~(b)(c)(d) & 9.8002 & 3.18 \\
SDSS 102700$-$010424 & 0.34379 & (a)(b)(c)(d) & 9.6652 & 3.00 \\
SDSS 102920$-$004747* & 0.25854 & (a)(b)~~~~(d) & 9.6349 & 3.15 \\
SDSS 104132$-$005057 & 0.30281 & (a)(b)(c)(d) & 9.6918 & 3.24 \\
SDSS 104431$-$001118 & 0.55928 & ~~~~(b)(c)(d) & 9.6604 & 3.10 \\
SDSS 105151$-$005117* & 0.35892 & (a)(b)(c)(d) & 9.6571 & 2.99 \\
SDSS 105228$-$010448 & 0.43549 & ~~~~(b)(c)(d) & 9.6868 & 2.98 \\
SDSS 105337+005958 & 0.47640 & ~~~~(b)(c)(d) & 9.6817 & 2.92 \\
SDSS 105706$-$004145 & 0.18776 & ~~~~(b)(c)(d) & 9.7255 & 2.81 \\
SDSS 111231$-$002534 & 0.54355 & ~~~~(b)(c)(d) & 9.6981 & 3.04 \\
SDSS 111353$-$000217 & 0.44544 & (a)(b)(c)(d) & 9.6723 & 2.99 \\
SDSS 114510+011056 & 0.62529 & (a)(b)~~~~(d) & 9.7346 & 3.22 \\
SDSS 115758$-$002220 & 0.25960 & (a)(b)(c)(d) & 9.6684 & 2.99 \\
SDSS 122004$-$002539* & 0.42118 & ~~~~(b)(c)(d) & 9.6690 & 2.98 \\
SDSS 123209+005015* & 0.47652 & (a)(b)~~~~(d) & 9.7880 & 2.55 \\
SDSS 130002$-$010601 & 0.30732 & (a)(b)(c)(d) & 9.6648 & 3.02 \\
SDSS 130713$-$003601 & 0.17012 & (a)(b)(c)(d) & 9.6627 & 2.96 \\
SDSS 132748$-$001021* & 0.47933 & ~~~~(b)(c)(d) & 9.6853 & 2.93 \\
SDSS 134113$-$005315* & 0.23745 & (a)(b)(c)(d) & 9.7190 & 2.91 \\
SDSS 134459$-$001559* & 0.24475 & (a)(b)(c)(d) & 9.6579 & 3.01 \\
SDSS 134507$-$001900 & 0.41867 & (a)(b)(c)(d) & 9.6618 & 3.05 \\
SDSS 135553+001137 & 0.45959 & (a)(b)(c)(d) & 9.6902 & 3.05 \\
SDSS 140827+004815 & 0.44221 & ~~~~(b)(c)(d) & 9.6605 & 3.16 \\
SDSS 141637+003352* & 0.43367 & (a)~~~~(c)(d) & 9.6814 & 2.99 \\
SDSS 142648+005323 & 0.21956 & (a)(b)(c)(d) & 9.6572 & 2.96 \\
SDSS 145221+002359 & 0.45792 & (a)(b)(c)(d) & 9.6263 & 3.13 \\
SDSS 150629+003543 & 0.36980 & (a)(b)(c)(d) & 9.7086 & 2.91 \\
SDSS 152110+000304 & 0.46541 & (a)~~~~(c)(d) & 9.6417 & 3.08 \\
SDSS 153306+000635 & 0.58916 & (a)~~~~(c)(d) & 9.6755 & 2.96 \\
SDSS 165627+623226 & 0.18477 & (a)~~~~(c)(d) & 9.6071 & 2.91 \\
SDSS 170441+604430* & 0.37124 & ~~~~(b)(c)(d) & 9.8926 & 4.31 \\
SDSS 170956+573225 & 0.52167 & (a)(b)(c)~~~~ & 9.6970 & 2.97 \\
SDSS 171441+644155* & 0.28478 & (a)(b)(c)~~~~ & 9.7178 & 3.19 \\
SDSS 172032+551330 & 0.27235 & (a)(b)(c)(d) & 9.6735 & 2.95 \\
SDSS 172059+612811 & 0.23619 & ~~~~(b)(c)(d) & 9.6507 & 3.20 \\
SDSS 172206+565451 & 0.42554 & (a)(b)(c)(d) & 9.7407 & 2.78 \\
SDSS 172446+575453 & 0.67647 & (a)(b)~~~~(d) & 9.6256 & 3.27 \\
SDSS 172543+580604 & 0.29197 & ~~~~(b)(c)(d) & 9.6598 & 2.92 \\
SDSS 172554+562458 & 0.73570 & (a)(b)~~~~(d) & 9.5996 & 2.58 \\
SDSS 173311+535457 & 0.30721 & (a)(b)(c)(d) & 9.6678 & 3.25 \\
SDSS 173602+554040 & 0.49675 & (a)(b)(c)(d) & 9.6677 & 2.83 \\
SDSS 173638+535432 & 0.40808 & (a)(b)(c)(d) & 9.6750 & 3.20 \\
SDSS 173721+550321 & 0.33299 & (a)(b)~~~~(d) & 9.6870 & 2.74 \\
SDSS 232640$-$003041 & 0.58191 & ~~~~(b)(c)(d) & 9.6721 & 3.14 \\
SDSS 232801+001705 & 0.41115 & ~~~~(b)(c)(d) & 9.6433 & 3.11 \\
SDSS 233517+010307 & 0.62401 & (a)(b)~~~~(d) & 9.7110 & 2.69 \\
SDSS 234145$-$004640 & 0.52386 & (a)(b)(c)(d) & 9.6914 & 3.04 \\
SDSS 235156$-$010913* & 0.17407 & ~~~~(b)(c)(d) & 9.7148 & 2.92 \\
SDSS 235439+005751 & 0.38974 & (a)(b)(c)(d) & 9.6451 & 2.84 \\
SDSS 235441$-$000448 & 0.27877 & (a)(b)(c)(d) & 9.6000 & 2.72 \\
SDSS 235732$-$002845 & 0.50829 & ~~~~(b)(c)(d) & 9.6929 & 2.91 \\

\enddata
\end{deluxetable}

\begin{deluxetable}{l@{\extracolsep{40pt}}c}
\tablewidth{0pt} \tablecaption{\baselineskip=12pt Alternative names. The table lists the SDSS name and an
alternative catalog name for the 95 SDSS quasars that passed at least three of the four standard tests
described in \S~\ref{sec:tests} and  that have been previously listed in other
catalogs.\label{tab:alternatives}} \tablehead{ \colhead{SDSS Name} &\colhead{Alternative Name}}
\startdata
SDSS 001327+005231 & LBQS 0010+0035 \\
SDSS 005905+000651 & LBQS 0056$-$0009 \\
SDSS 013352+011345 & UM 338 \\
SDSS 015950+002340 & MRK 1014 \\
SDSS 024240+005727 & E 0240+007 \\
SDSS 024954+010148 & US 3213 \\
SDSS 025007+002525 & LEDA 138544 \\
SDSS 025432$-$004220 & LBQS 0251$-$0054 \\
SDSS 030048+005440 & US 3513 \\
SDSS 030313$-$001457 & $[$HB89$]$ 0300$-$004 \\
SDSS 095625$-$000353 & 2QZ J095625.8$-$000354 \\
SDSS 102502+003126 & LBQS 1022+0046 \\
SDSS 102920$-$004747 & LBQS 1026$-$0032 \\
SDSS 105151$-$005117 & PG 1049$-$005 \\
SDSS 122004$-$002539 & 2QZ J122004.3$-$002540 \\
SDSS 123209+005015 & LBQS 1229+0106 \\
SDSS 132748$-$001021 & 2QZ J132748.1$-$001021 \\
SDSS 134113$-$005315 & LBQS 1338$-$0038 \\
SDSS 134459$-$001559 & LBQS 1342$-$0000 \\
SDSS 141637+003352 & 2QZ J141637.4+003351 \\
SDSS 170441+604430 & $[$HB89$]$ 1704+608 \\
SDSS 171441+644155 & HS 1714+6445 \\
SDSS 235156$-$010913 & $[$HB89$]$ 2349$-$014 \\
\enddata
\end{deluxetable}
\clearpage

\section{Analysis of the SDSS Data Release One Sample.}
\label{appendix:C}

In this Appendix, we present additional results for quasars from the SDSS Data Release One (DR1,
Schneider et al. 2003)\nocite{DR1}. We selected quasars spectra from the DR1 sample using the procedure
that we described in \S~\ref{subsubsec:spectra} and applied earlier to the EDR sample. There are 3429
quasar spectra in the DR1 sample (702 in the EDR sample) that potentially have measurable O~{\sc iii}
emission lines, of which 1700 (260 in EDR) have lines suitable for precision mission (i. e., pass the
'preliminary filtering' described in \S~\ref{subsubsec:spectra}).  A total of 431 quasars in the DR1
sample (105 in the EDR sample) pass three of the four standard test described in \S~\ref{sec:tests}.
Finally, 165 quasars in the DR1 sample (42 in the EDR sample) pass all four of the standard tests.

The results for the DR1 sample are consistent with the results for the SDSS Early Data Release shown in
Table \ref{tab:average} and Table \ref{tab:linearfit} of the main text of this paper, but the DR1 results
have smaller errors as expected from the larger data sample. For example, the standard sample of the DR1
data release has approximately four times as many quasar spectra as the EDR standard sample (162 versus
42) and, as expected, the error on the measurement of $\left<\alpha/\alpha(0)\right>$ is about a factor
of two smaller for the DR1 standard sample than for the EDR standard sample.

All but five of the EDR standard sample of quasars are included in the DR1 standard sample. If we add
these five EDR quasars to the 165 DR1 quasars, we obtained a combined EDR plus DR1 standard sample of 170
quasars for which

\begin{equation}
\frac{\left<\alpha(z)\right>}{\alpha(0)} ~=~ 1.00011 \pm 0.00007 \, . \label{eq:averageunion}
\end{equation}
This result is essentially identical to the result found for the DR1 sample alone (see the first row of
Table~\ref{tab:DR1average}.

\begin{table}[!ht]
\centering \caption[]{\baselineskip=12pt This table presents the measured weighted average value of
$\alpha/\alpha(0)$ for the SDSS Data Release One sample as well as the measured average for 17
alternative cuts, defined in $\S~\ref{subsec:alternatives}$, on the data.  The number of quasar spectra
that pass the cuts defining each sample is given in the second column.  No alternative sub-ample of the
DR1 sample produces an average value of $\alpha$ significantly different from the value obtained for the
standard sample of the DR1.  For the different sub-samples, the values obtained for the DR1 are also in
agreement with the values obtained for the EDR SDSS data release.} \label{tab:DR1average}
\begin{tabular}{ccc}
\tableline\tableline
Sample & Sample Size& Average $\alpha/\alpha(0)$\\
\tableline
Standard       & 165&     $1.00012 \pm 0.00007$\\
Remove bad $z$   & 162&     $1.00012 \pm 0.00007$\\
Unweighted errors    & 165&     $1.00023 \pm 0.00014$\\
Strict H$\beta$ limit      & 128&    $ 1.00002 \pm 0.00007$\\
Signal-to-noise of 10:1  & 216&     $1.00013 \pm 0.00007$\\
11 point KS test    & 290&    $ 1.00016 \pm 0.00006$\\
EW not area            & 160&     $1.00013 \pm 0.00007$\\
$\chi^2$ instead of KS           & 140&    $ 1.00012 \pm 0.00008$\\
Omit KS test & 308&     $1.00015 \pm 0.00006$\\
Omit peak line up& 205&    $ 1.00012 \pm 0.00007$\\
Omit H$\beta$ test& 180&    $ 1.00016 \pm 0.00007$\\
Omit signal-to-noise test& 233&    $ 1.00014 \pm 0.00007$\\
Remove worst outlier& 164&     $1.00012 \pm 0.00007$\\
$z <.3632$ &  89&     $1.00012 \pm 0.00010$\\
$z >.3632$  &  76&     $1.00012 \pm 0.00011$\\
$\Omega_m=1, \Omega_\Lambda=0$ & 165&    $ 1.00012 \pm 0.00007$\\
$\Omega_m=0, \Omega_\Lambda=1$ & 165&    $ 1.00012 \pm 0.00007$\\
Add $R(0)$ and $\sigma_{0}$ & 166&    $ 1.00004 \pm 0.00007$\\
Add $R(0)$ and $0.1\sigma_{0}$ & 166&    $ 1.00004 \pm 0.00007$\\
\tableline
\end{tabular}
\end{table}

\begin{table}[!hbt]
\centering \caption[]{\baselineskip=12pt Best linear fit for the DR1 sample.  Here $\alpha^2(t) =
\alpha^2_{\rm fit, 0}[1 + H_0 S t]$, where the slope $S$ is defined by equation~(\ref{eq:defnslope}). In
the table, the value of $\alpha^2_{\rm fit, 0}/\alpha^2 ({\rm local ~ meas.})$ is defined by
equation~(\ref{eq:ratioofalphas}). The time $t$ is calculated from equation~(\ref{eq:defntime}) for a
universe with the present composition of $\Omega_m=0.3,\Omega_\Lambda=0.7$.  The results presented here
for the DR1 sample are consistent with  those presented previously in Table \ref{tab:linearfit} for the
EDR sample, but the DR1 results are more are more accurate. All of the sub-samples of the DR1 give
consistent results.} \label{tab:DR1slope}
\begin{tabular}{cccc}
\tableline\tableline
Sample & Sample Size &$\alpha^2_{\rm fit, 0}/\alpha^2 ({\rm local ~ meas.})$& Slope $S$ \\
\tableline
Standard       & 165&                      $ 1.0008 \pm 0.0004$&      $ -0.0021 \pm 0.0016$\\
Unweighted errors& 165&                    $ 1.0007 \pm 0.0010$&      $ -0.0008 \pm 0.0030$\\
Strict H$\beta$  & 128&                    $ 1.0003 \pm 0.0004$&      $ -0.0009 \pm 0.0015$\\
Signal-to-noise of 10:1       & 216&       $ 1.0007 \pm 0.0004$&      $ -0.0017 \pm 0.0016$\\
11 point KS test& 290&                     $ 1.0011 \pm 0.0004$&      $ -0.0029 \pm 0.0015$\\
EW not area& 160&                          $ 1.0007 \pm 0.0004$&      $ -0.0017 \pm 0.0016$\\
$\chi^2$ instead of KS & 140&              $ 1.0009 \pm 0.0006$&      $ -0.0027 \pm 0.0021$\\
Omit KS test & 308&                        $ 1.0011 \pm 0.0004$&      $ -0.0030 \pm 0.0015$\\
Omit peak line up& 205&                    $ 1.0005 \pm 0.0004$&      $ -0.0008 \pm 0.0014$\\
Omit H$\beta$ test& 180&                   $ 1.0009 \pm 0.0004$&      $ -0.0023 \pm 0.0016$\\
Omit signal-to-noise test& 23 &            $ 1.0007 \pm 0.0004$&      $ -0.0017 \pm 0.0016$\\
Remaove worst outlier& 164&                $ 1.0008 \pm 0.0004$&      $ -0.0022 \pm 0.0016$\\
$z <.3632$ &  89&                          $ 1.0011 \pm 0.0008$&      $ -0.0040 \pm 0.0039$\\
$z >.3632$ &  76&                          $ 1.0034 \pm 0.0017$&      $ -0.0094 \pm 0.0049$\\
$\Omega_m=1, \Omega_\Lambda=0$  & 165&$ 1.0006 \pm 0.0003$&      $ -0.0007 \pm 0.0005$\\
$\Omega_m=0, \Omega_\Lambda=1$& 165&       $ 1.0007 \pm 0.0004$&      $ -0.0018 \pm 0.0013$\\
Add $R(0)$ and $\sigma_{0}$ & 166&         $ 1.0003 \pm 0.0005$&      $ -0.0003 \pm 0.0017$\\
Add $R(0)$ and $0.1\sigma_{0}$ & 166&      $ 1.0003 \pm 0.0005$&      $ -0.0003 \pm 0.0017$\\
\tableline
\end{tabular}
\end{table}

\end{document}